\documentclass[journal=jacsat,manuscript=article]{achemso}
\setkeys{acs}{articletitle=true}

\usepackage[version=3]{mhchem} 
\usepackage[]{mathrsfs}
\usepackage[]{listings}
\usepackage[latin1]{inputenc}
\usepackage{amssymb}
\usepackage{epsfig}
\usepackage{xspace}
\usepackage{pstricks}
\usepackage{psfrag}
\usepackage{textgreek}
\usepackage{color}
\usepackage{amsmath}

\newcounter{myitem}

\newcommand{\resetmycnt}{\setcounter{myitem}{1}}

\resetmycnt

\newcommand{\figdir}  {./figures/paper}
\newcommand{\figdirSI}{./figures/SI}

\author{\vspace*{-0.1cm}\it Shaokai Zhou}
\affiliation{Laboratory of 2D Optoelectronics and Nanoelectronics (L2DON), State Key Laboratory of Quantum Functional Materials, Department of Materials Science and Engineering, Southern University of Science and Technology, 1088 Xueyuan Blvd, 518055 Shenzhen, China}
\alsoaffiliation{These authors contributed equally to this work.}
\author{\vspace*{-0.1cm}\it Haihui Cai}
\affiliation{Laboratory of 2D Optoelectronics and Nanoelectronics (L2DON), State Key Laboratory of Quantum Functional Materials, Department of Materials Science and Engineering, Southern University of Science and Technology, 1088 Xueyuan Blvd, 518055 Shenzhen, China}
\alsoaffiliation{These authors contributed equally to this work.}
\author{\vspace*{-0.1cm}\it Yehao Wu}
\affiliation{Laboratory of 2D Optoelectronics and Nanoelectronics (L2DON), State Key Laboratory of Quantum Functional Materials, Department of Materials Science and Engineering, Southern University of Science and Technology, 1088 Xueyuan Blvd, 518055 Shenzhen, China}
\alsoaffiliation{These authors contributed equally to this work.}
\author{\vspace*{-0.1cm}\it Yufeng Min}
\affiliation{School of Microelectronics, University of Science and Technology of China, 230026 Hefei, China}
\author{\vspace*{-0.1cm}\it Renchen Yuan}
\affiliation{School of Microelectronics, University of Science and Technology of China, 230026 Hefei, China}
\author{\vspace*{-0.1cm}\it Yezhu Lv}
\affiliation{Laboratory of 2D Optoelectronics and Nanoelectronics (L2DON), State Key Laboratory of Quantum Functional Materials, Department of Materials Science and Engineering, Southern University of Science and Technology, 1088 Xueyuan Blvd, 518055 Shenzhen, China}
\author{\vspace*{-0.1cm}\it Jianming Huang}
\affiliation{Laboratory of 2D Optoelectronics and Nanoelectronics (L2DON), State Key Laboratory of Quantum Functional Materials, Department of Materials Science and Engineering, Southern University of Science and Technology, 1088 Xueyuan Blvd, 518055 Shenzhen, China}
\author{\vspace*{-0.1cm}\it Yuanyuan Shi}
\affiliation{School of Microelectronics, University of Science and Technology of China, 230026 Hefei, China}
\email{yuanyuanshi@ustc.edu.cn}
\author{\vspace*{-0.1cm}\it Yury Yuryevich Illarionov}
\affiliation{Laboratory of 2D Optoelectronics and Nanoelectronics (L2DON), State Key Laboratory of Quantum Functional Materials, Department of Materials Science and Engineering, Southern University of Science and Technology, 1088 Xueyuan Blvd, 518055 Shenzhen, China}
\email{illarionov@sustech.edu.cn}

\title[An \textsf{achemso} demo]
  {Mobile charges in MoS$_2$/high-k oxide transistors: from abnormal instabilities to memory-like dynamics}

\begin{document}



\begin{abstract}
MoS$_2$ field-effect transistors (FETs) with high-\textit{k} oxides currently lag behind silicon standards in bias and temperature stability due to ubiquitous border oxide traps that cause clockwise (CW) hysteresis in gate transfer characteristics. While suppressing this effect is typically mandatory for logic FETs, here we explore an alternative strategy where the initial CW hysteresis can be dynamically overcome by stronger counterclockwise (CCW) hysteresis towards memory-like dynamics. We systematically compare hysteresis in similar back-gated MoS$_2$/HfO$_2$ and MoS$_2$/Al$_2$O$_3$ FETs up to 275\textdegree C. At room temperature, both devices initially show sizable CW hysteresis. However, at 175\textdegree C MoS$_2$/HfO$_2$ FETs exhibit dominant CCW dynamics coupled with self-doping and negative differential resistance (NDR) effects. Our compact model suggests that this behavior is caused by the drift of mobile oxygen vacancies (\textit{V}\({}_{\mathrm{O}}^{+}\) or \textit{V}\({}_{\mathrm{O}}^{2+}\)) within HfO$_2$ which also causes negative $V_{\mathrm{th}}$ shift under a constant positive bias stress. This alternative mechanism effectively overrides the initial CW hysteresis and enables intrinsic memory functionality that can be enhanced by using narrower gate bias sweep ranges. In contrast, the MoS$_2$/Al$_2$O$_3$ FETs display only minor CCW dynamics even at 275\textdegree C due to higher drift activation energies for the same vacancies, thereby maintaining superior stability. Our results reveal an insulators selection paradigm: Al$_2$O$_3$ layers are better suited to suppress detrimental negative $V_{\mathrm{th}}$ shifts in MoS$_2$ logic FETs at high temperatures, whereas their HfO$_2$ counterparts can serve as active memory layers that would exploit these abnormal instabilities.

\end{abstract}

\clearpage

Moore's scaling of silicon-based FETs is approaching its fundamental limits due to sizable short-channel effects and mobility degradation in thin Si nanosheets~\cite{UCHIDA02}. This creates an urgent need for alternative channel materials such as 2D semiconductors~\cite{LEMME22,GHOSH25} which could replace or substitute Si in next-generation integrated circuits~\cite{SCHWIERZ15,DAS21,LEMME22}. Atomically thin MoS$_2$ that enables superior electrostatic control and offers reasonable compatibility with conventional high-k oxides such as HfO$_2$ and Al$_2$O$_3$ has emerged as a leading candidate for 2D n-FETs~\cite{CHUNG24,MORTELMANS24,GHOSH25}. Furthermore, recent research advances~\cite{WACHTER17,LEMME22,HUANG22,TAN23,PENDURTHI24} have made it possible to move forward from lab prototypes to the trial integration of the MoS$_2$ FETs with HfO$_2$ and Al$_2$O$_3$ into the industrial process lines~\cite{ASSELBERGHS20,DOROW22,CHUNG24,JAYACHANDRAN24,MORTELMANS24}. However, these devices still face severe stability and reliability limitations caused by various defects in gate insulators, with border oxide traps~\cite{FLEETWOOD92,ILLARIONOV20A} and mobile charges~\cite{KNOBLOCH23,PROVIAS23} being the most ubiquitous. As a result, the pathway of MoS$_2$ FETs and other 2D devices to the mass-production is delayed since it is still hard to compete with Si technologies in stable long-term operation. 

A powerful diagnostic tool to benchmark stable operation of MoS$_2$ FETs can be offered by comprehensive analysis of time-dependent hysteresis dynamics in the gate transfer ($I_{\mathrm{D}}$-$V_{\mathrm{G}}$) characteristics~\cite{ILLARIONOV16A,ILLARIONOV20A,KNOBLOCH23,KARL25}. Hysteresis serves as a sensitive indicator for the stability of MoS$_2$ FETs and can also reveal valuable physical phenomena that could be exploited for beyond-FET applications. That is because it directly reflects threshold voltage fluctuations and captures key information about dynamic processes such as charge trapping by oxide defects and drift of mobile charges~\cite{PROVIAS23}. However, many existing studies on 2D FETs miss this opportunity to get in-depth information and use just a single sweep at a random sweep rate which only allows to speculate that hysteresis is ``negligible'' or ``near-zero''~\cite{ROH16,VU18,CHO21E,VENKA24,VENKA24,FAN25,LAN25} with no relation to the real physical picture. As a result, no systematic comparison of time- and temperature-dependent hysteresis dynamics in MoS$_2$ FETs with HfO$_2$ and Al$_2$O$_3$ insulators under identical conditions can be found in the literature. This prevents clear understanding of intrinsic stability concerns introduced by these two most widely used high-k gate oxides and thus impedes formulation of targeted optimization strategies that could finally enable MoS$_2$/high-k FETs with competitive stability. Furthermore, the lack of detailed studies focused specifically on abnormal CCW dynamics impedes possible ways for the practical use of this phenomena in memory devices as an alternative to suppression.  

Here we perform a systematic comparison of hysteresis dynamics in MoS$_2$/HfO$_2$ and MoS$_2$/Al$_2$O$_3$ FETs fabricated using identical methods at sweep times up to tens of kiloseconds and temperatures varied from 25\textdegree C to 275\textdegree C and support our findings with bias stress measurements. Our results reveal that at room temperature both devices exhibit a CW hysteresis that is caused by charge trapping and thus defined by energetic alignment of oxide defect bands. However, at 175\textdegree C our MoS$_2$/HfO$_2$ FETs exhibit change of hysteresis to the CCW direction coming together with NDR and self-doping effects. Our qualitative compact model for mobile charges in oxides nicely reproduces the CCW hysteresis with all related phenomena and suggests the drift of positive oxygen vacancies in HfO$_2$ as the primary reason that may also result in controllable memory-like dynamics. However, in Al$_2$O$_3$ insulators similar trends are barely visible even at the temperatures as high as 275\textdegree C due to larger migration barriers of the same vacancies, thus making this insulator more suitable to mitigate detrimental negative shifts of $V_{\mathrm{th}}$ under the positive bias stress.

\section{Results and Discussion}

\subsection{Investigated devices and mesurement technique}

\begin{figure}[]
\vspace{0mm}
\begin{minipage} {\textwidth} 
\hspace{1.2cm}
  \includegraphics[width=13cm]{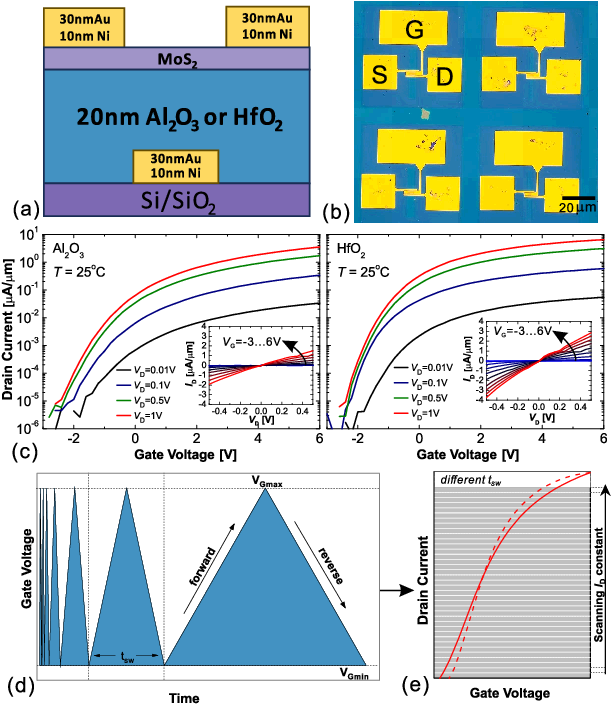} 
\caption{\label{Fig.1} (a) Schematic layout of our back-gated MoS$_2$/HfO$_2$ and MoS$_2$/Al$_2$O$_3$ FETs fabricated using the same process. The channel is made of CVD-grown MoS$_2$ films taken from the same batch. (b) Optical image of four devices inside the array containing tens of FETs. (c) $I_{\mathrm{D}}$-$V_{\mathrm{G}}$ characteristics of MoS$_2$ FETs with Al$_2$O$_3$ (left) and HfO$_2$ (right) measured at different $V_{\mathrm{D}}$. The insets show the corresponding $I_{\mathrm{D}}$-$V_{\mathrm{D}}$ curves.  (d) Schematics of the subsequent $V_{\mathrm{G}}$ sweeps with increased $t_{\mathrm{sw}}$ used in our hysteresis measurements. In some measurements maximum $t_{\mathrm{sw}}$ that we have reached was above 10$\,$ks. (e) Schematics of our universal mapping method~\cite{LV25M} which suggests scanning the $I_{\mathrm{D}}$ values to obtain minimum (lower UHF) and maximum (upper UHF) from the extracted family of $\Delta V_{\mathrm{H}}$(1/$t_{\mathrm{sw}}$) curves.}
\end{minipage}
\end{figure}

We examine back-gated MoS$_2$ FETs with HfO$_2$ and Al$_2$O$_3$ gate insulators with the schematic layout shown in Fig.1a. To enable consistent comparison of the device stability, for both types of FETs we used single-layer MoS$_2$ films grown by chemical-vapor deposition (CVD) taken from the same batch and also employed the identical fabrication processes. First, Ni(10$\,$nm)/Au(30$\,$nm) local back gate electrodes were created on a pretreated Si/SiO$_2$ substrate via photolithography and e-beam evaporation. Then a 20$\,$nm-thick HfO$_2$ or Al$_2$O$_3$ insulator was grown by atomic layer deposition (ALD) at 250\textdegree C. Next, the MoS$_2$ film was transferred through a wet transfer process, followed by residual removal and patterning of the channel via photolithography and reactive ion etching. Finally, the source/drain electrodes were defined by photolithography and metalized with Ni(10$\,$nm)/Au(30$\,$nm) via e-beam evaporation and lift-off process. The obtained arrays with many tens of MoS$_2$ FETs of both types (Fig.1b) with channel length $L$ and width $W$ varied between 5 and 100$\,\mu$m allowed us to perform systematic measurements while verifying the reproducibility of all observed trends.     

The electrical characterization of our MoS$_2$ FETs was performed in vacuum and in complete darkness. First we have checked the $I_{\mathrm{D}}$-$V_{\mathrm{G}}$ characteristics at different drain voltages $V_{\mathrm{D}}$. As shown in Fig.1c, for both MoS$_2$/Al$_2$O$_3$ and MoS$_2$/HfO$_2$ FETs decent performance with on/off current ratios of about 10$^6$ is obtained, while the output ($I_{\mathrm{D}}$-$V_{\mathrm{D}}$) curves also look reasonable. Next we measured the hysteresis using subsequent double sweeps of the  $I_{\mathrm{D}}$-$V_{\mathrm{G}}$ characteristics at $V_{\mathrm{D}}\,=\,0.1\,$V with sweep times $t_{\mathrm{sw}}$ increased from tens of seconds to over 10$\,$ks (Fig.1d), while repeating the measurements with different $V_{\mathrm{G}}$ sweep ranges. For consistent extraction of the hysteresis width $\Delta V_{\mathrm{H}}$ vs. 1/$t_{\mathrm{sw}}$ dependences we used our standardized universal mapping method~\cite{LV25M}. This approach suggests scanning the constant current value to extract the lower and upper universal hysteresis functions (UHFs)~\cite{LV25M}, as schematically shown in Fig.1e. In contrast to the conventional constant current approach used in our previous studies~\cite{ILLARIONOV17A,KNOBLOCH23}, the mapping method captures complex hysteresis dynamics like the CW/CCW switching caused by two competing mechanisms such as charge trapping and drift of mobile charges observed in this work.  

\subsection{Hysteresis dynamics at room temperature}

\begin{figure}[!h]
\vspace{0mm}
\begin{minipage} {\textwidth} 
\hspace{0.1cm}
  \includegraphics[width=16.0cm]{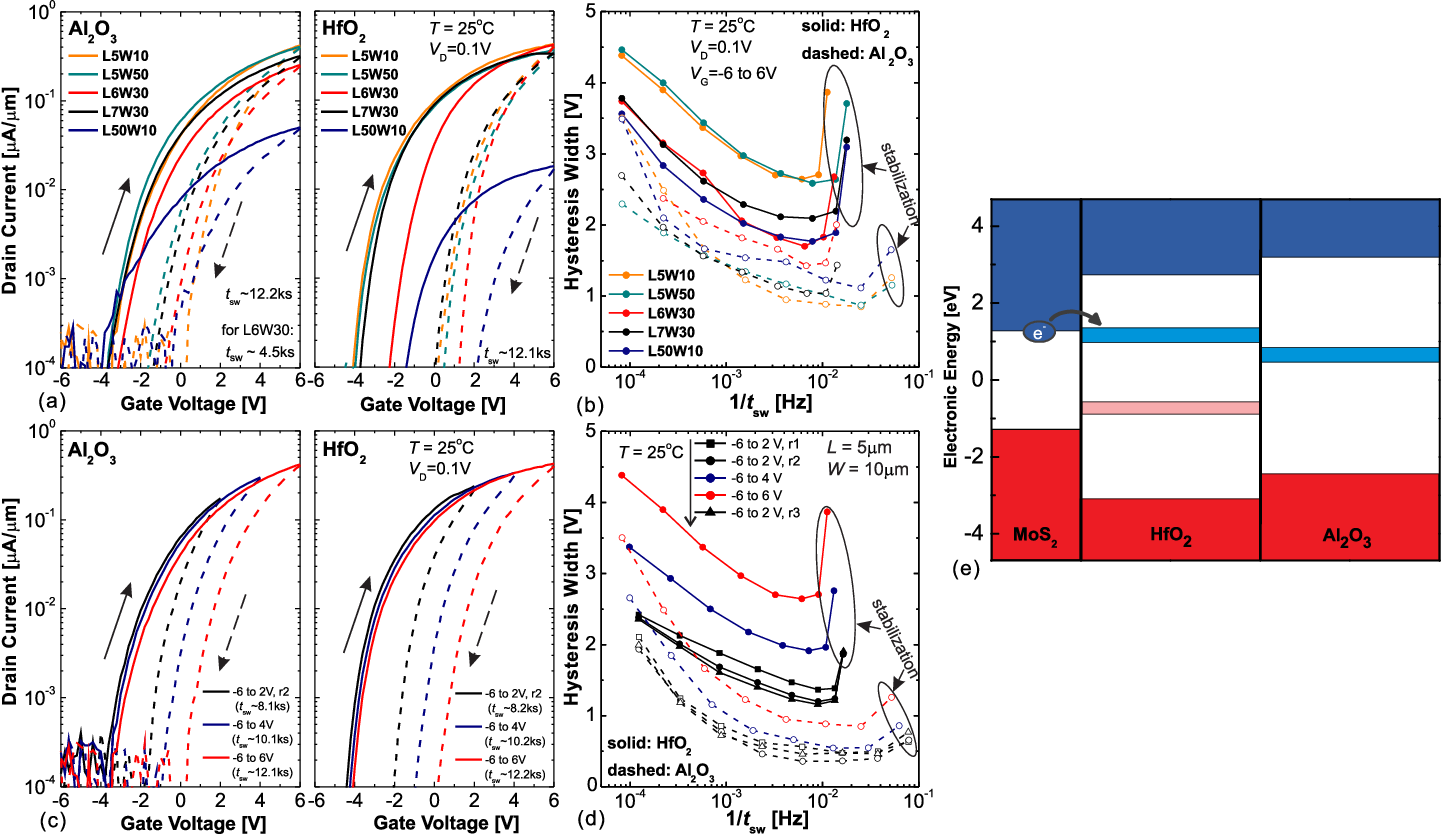} 
\caption{\label{Fig.2} (a) Double sweep $I_{\mathrm{D}}$-$V_{\mathrm{G}}$ characteristics of 5 MoS$_2$ FETs with Al$_2$O$_3$ (left) and HfO$_2$ (right) with different $L$ and $W$ measured using the slowest achieved sweeps at 25\textdegree C. A purely CW hysteresis is present for all devices. (b) The corresponding $\Delta V_{\mathrm{H}}$ vs. $1/t_{\text{sw}}$ dependences showing that while there is some device-to-device variability for both types of devices, in overall the CW hysteresis is smaller for the devices with Al$_2$O$_3$. (c) Double sweep $I_{\mathrm{D}}$-$V_{\mathrm{G}}$ characteristics of the representative MoS$_2$ FETs with Al$_2$O$_3$ (left) and HfO$_2$ (right) measured using the slowest $t_{\mathrm{sw}}$ and different $V_{\mathrm{G}}$ sweep ranges. (d) The corresponding $\Delta V_{\mathrm{H}}$ vs. $1/t_{\text{sw}}$ curves confirm that for the MoS$_2$/Al$_2$O$_3$ FETs hysteresis is smaller for all $V_{\mathrm{G}}$ sweep ranges. Remarkably, the results for -6 to 2$\,$V sweep range repeated in three rounds (r1, r2, r3) in the beginning and in the end of experiment (as illustrated by the arrow next to the legend) show perfect reproducibility. (e) Band diagram showing relative energetic alignments between the conduction band edge $E_{\mathrm{C}}$ of the MoS$_2$ n-channel and known fundamental defect bands in HfO$_2$ and Al$_2$O$_3$~\cite{RZEPA18,ILLARIONOV20A}. The upper defect band of HfO$_2$ is closer to $E_{\mathrm{C}}$ of MoS$_2$, while the defect band of Al$_2$O$_3$ is situated deeper. This explains smaller CW hysteresis in the latter case, since deeper traps in Al$_2$O$_3$ get activated at slower sweeps. }
\end{minipage}
\end{figure}

First we reveal the impact of gate insulator and device-to-device variability on the hysteresis dynamics at room temperature by examining MoS$_2$ FETs with different $L$ and $W$. In Fig.2a we show the double sweep $I_{\mathrm{D}}$-$V_{\mathrm{G}}$ characteristics measured for 5 MoS$_2$/Al$_2$O$_3$ and 5 MoS$_2$/HfO$_2$ devices using the slowest achieved $t_{\mathrm{sw}}$ and the $V_{\mathrm{G}}$ sweep range of -6 to 6$\,$V. A purely CW hysteresis is observed in all cases. Then we use our full measurement datasets consisting of 8 sweeps with $t_{\mathrm{sw}}$ ranging between tens of seconds and over ten kiloseconds to perform the full mapping of hysteresis dynamics. As shown in Fig.S1 in the Supplementary Information (SI), for both types of our devices only CW hysteresis is present within the whole range of $t_{\mathrm{sw}}$ and thus the $\Delta V_{\mathrm{H}}$(1/$t_{\mathrm{sw}}$) dependences can be consistently expressed with the upper UHFs. In Fig.2b we show the resulting $\Delta V_{\mathrm{H}}$(1/$t_{\mathrm{sw}}$) curves for all 10 devices. While some variability is present within 5 devices of each type, there is no obvious correlation between the hysteresis magnitude and the channel dimensions. Furthermore, from Fig.2b we can reliably conclude that the CW hysteresis in MoS$_2$ FETs with Al$_2$O$_3$ is generally smaller than in their counterparts with HfO$_2$. To get more insights into this difference, we next select representative MoS$_2$ FETs of both types and measure the hysteresis using different $V_{\mathrm{G}}$ sweep ranges. The double sweep $I_{\mathrm{D}}$-$V_{\mathrm{G}}$ characteristics measured at the slowest $t_{\mathrm{sw}}$ and the corresponding $\Delta V_{\mathrm{H}}$(1/$t_{\mathrm{sw}}$) dependences are shown in Fig.2c,d, respectively. We can see that for both types of our MoS$_2$ FETs the CW hysteresis becomes larger for wider $V_{\mathrm{G}}$ sweep ranges and also increases for slower sweeps that is the typical feature of charge trapping by oxide defects~\cite{ILLARIONOV20A}. At the same time, it is again reconfirmed that for the MoS$_2$/Al$_2$O$_3$ FETs the CW hysteresis is generally smaller and starts to increase at considerably lower sweep frequencies for all three sweep ranges used. Remarkably, the results for the narrowest $V_{\mathrm{G}}$ sweep range of -6 to 2$\,$V measured before and after the use of wider $V_{\mathrm{G}}$ sweep ranges are nicely reproducible. However, stabilizing behavior of the devices which appears as a larger hysteresis measured for the first sweeps of each round has to be noted in all cases.   

The observed difference in room temperature hysteresis dynamics between MoS$_2$ FETs with Al$_2$O$_3$ and HfO$_2$ gate insulators goes in line with the energetic alignment of oxide defect bands with respect to the conduction band edge $E_{\mathrm{C}}$ of MoS$_2$ as shown in Fig.2e. These defect bands have fundamental positions inside the bandgap of high-k oxides which are relatively well known but may slightly depend on the oxide quality and growth conditions~\cite{RZEPA18,ILLARIONOV20A}. Namely, the upper defect band of HfO$_2$ is close to $E_{\mathrm{C}}$ of MoS$_2$. This facilitates trapping of electrons from the MoS$_2$ channel by HfO$_2$ defects~\cite{KNOBLOCH22} which explains a sizable CW hysteresis observed in our MoS$_2$/HfO$_2$ FETs already at faster sweeps. In contrast, the defect band of Al$_2$O$_3$ is over 0.5$\,$eV below $E_{\mathrm{C}}$ of MoS$_2$, which drastically increases the energy barrier for charge trapping by oxide traps. That is why the CW hysteresis in our MoS$_2$/Al$_2$O$_3$ FETs is generally smaller as compared to their counterparts with HfO$_2$. Furthermore, $\Delta V_{\mathrm{H}}$ starts increasing at smaller sweep frequencies as these deeper traps in Al$_2$O$_3$ have larger capture/emission time constants and thus need more time to get activated.  

\subsection{{MoS$_2$/HfO$_2$ FETs at high temperatures: memory-like performance}}

\subsubsection{{Experimental classification of the CCW hysteresis dynamics}}

\begin{figure}[!h]
\vspace{0mm}
\begin{minipage} {\textwidth} 
\hspace{2cm}
  \includegraphics[width=11.1cm]{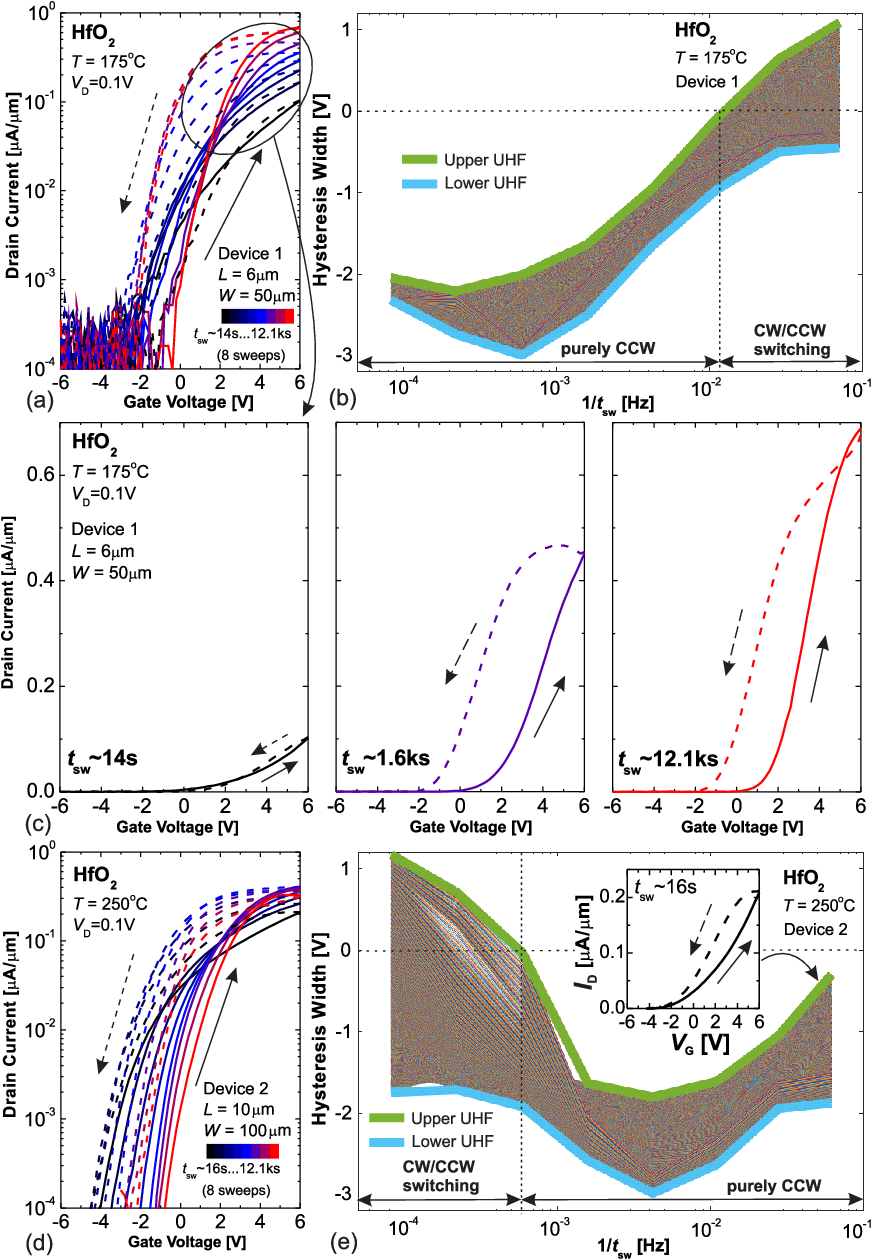} 
\caption{\label{Fig.3} (a) Double sweep $I_{\mathrm{D}}$-$V_{\mathrm{G}}$ characteristics of our MoS$_2$/HfO$_2$ FET measured at $T\,=\,$175\textdegree C using 8 subsequent sweeps with $t_{\mathrm{sw}}$ up to 12.1$\,$ks. The hysteresis dynamics changes from the CW/CCW switching at faster sweeps to the purely CCW hysteresis at slower sweeps. (b) The corresponding mapping results which clearly reveal the frequency ranges of both hysteresis dynamics and a distinct maximum of the CCW hysteresis. (c) Linear scale $I_{\mathrm{D}}$-$V_{\mathrm{G}}$ curves corresponding to (a) which show development of the CCW hysteresis, i.e. a weak effect at fast $t_{\mathrm{sw}}$, an NDR behavior at moderate $t_{\mathrm{sw}}$ and an increase of $I_{\mathrm{ON}}$ with localization and decay of the CCW hysteresis at slow $t_{\mathrm{sw}}$. (d,e) The related results measured at $T\,=\,$250\textdegree C. The CCW hysteresis maximum is shifted to faster sweeps and thus the CW/CCW switching is observed at slower sweeps. The inset in (e) highlights that NDR is not observed within the $t_{\mathrm{sw}}$ range used since mobile charges are too fast.}
\end{minipage}
\end{figure}

In Fig.3a we show the double sweep $I_{\mathrm{D}}$-$V_{\mathrm{G}}$ characteristics of our MoS$_2$/HfO$_2$ FET measured at $T\,=\,$175\textdegree C using 8 subsequent sweeps with $t_{\mathrm{sw}}$ up to 12.1$\,$ks. The corresponding hysteresis mapping results provided in Fig.3b reveal a totally different hysteresis behavior as compared to room temperature. Namely, the hysteresis changes from the CW/CCW switching at fast sweeps to the purely CCW one at slow sweeps. This is followed by sizable increase in the ON state current from one sweep to another (Fig.3a) which suggests progressive n-type doping of MoS$_2$ during the measurements. Remarkably, our full mapping results clearly separate the frequency regions in which the CW/CCW switching and purely CCW hysteresis are present, as in the former case the upper and lower UHFs are of the opposite signs and in the latter case they are both negative. At the same time, a distinct maximum of the CCW hysteresis is visible. This is in part similar to our previous observations for MoS$_2$/SiO$_2$ FETs~\cite{KNOBLOCH23} and hints at the contribution coming from mobile charges in the oxide. Furthermore, in Fig.3c we show the representative $I_{\mathrm{D}}$-$V_{\mathrm{G}}$ characteristics in linear scale and reveal that for the moderate $t_{\mathrm{sw}}$ an NDR effect is present during the reverse sweep. As will be discussed below, this memory-like behavior appears if positive mobile charges approaching the channel side of HfO$_2$ lag behind $V_{\mathrm{G}}$ changes near $V_{\mathrm{Gmax}}$. However, for slower $t_{\mathrm{sw}}$ a stronger increase of the ON current with no NDR effect is observed because the self-doping peaks at less positive $V_{\mathrm{G}}$ during the forward sweep, and thus by reaching $V_{\mathrm{Gmax}}$ the charges are already trapped at the channel side of oxide. The related results obtained at $T\,=\,$250\textdegree C (Fig.3d,e) show that the maximum of CCW hysteresis is shifted towards faster sweeps because of thermal activation of mobile charges. As a result, the CCW contribution becomes smaller at slow sweeps that results in the CW/CCW switching since the charge trapping contribution is still present. However, as the mobile charges are too fast at this temperature, we do not observe any sizable NDR effect. For instance, the inset of Fig.3e suggests that only some signs of the current plateau are visible in the beginning of reverse sweep even at the fastest $t_{\mathrm{sw}}$ used.

We also note that the CCW hysteresis dynamics starts to appear in most our MoS$_2$/HfO$_2$ FETs at $T\,=\,$125\textdegree C for slow $t_{\mathrm{sw}}$, with some outlying devices with likely more vacancies in the oxide showing NDR behavior already at this temperature. However, $T\,=\,$175\textdegree C appears to be the optimum for detailed analysis as the CCW hysteresis becomes dominant at this temperature. Furthermore, the observed CCW dynamics are well reproducible between different devices that simplifies analysis if certain devices get failed during the measurements. More detailed results on temperature dependence and reproducibility between different devices can be found in Fig.S2-S3 in the SI.

\subsubsection{{Understanding the observed trends via compact modeling}}

To reveal the physical origin of the complex high-temperature hysteresis dynamics in our MoS$_2$/HfO$_2$ FETs, we implement a compact model for the drift of mobile ions in the gate oxide while disregarding the CW contribution coming from oxide traps that is studied elsewhere by considering oxide defect bands~\cite{KNOBLOCH22,KNOBLOCH23}. The model incorporates thermally activated hopping of mobile charges with the diffusion coefficient given as 

\begin{equation}
D = D_0 \exp\left(-\frac{q E_{\mathrm{A}}}{k_{\mathrm{B}} T}\right)
\end{equation}
where $D_0$ is a pre-factor coefficient and $E_{\mathrm{A}}$ is the energy activation barrier for the migration of ions. Following the Einstein relation, this diffusion coefficient sets the mobility of ions and consequently their drift velocity under electric field given by the applied $V_{\mathrm{G}}$. As a result, the position of mobile charges in the oxide $x(t)$ can be calculated at any time point and used to elaborate the resulting time-dependent threshold voltage shift as

\begin{equation}
\Delta V_{\mathrm{th}}(t) = -\frac{Q_{\mathrm{max}}}{C_{\mathrm{ox}}} \left(1 - \frac{x(t)}{d_{\mathrm{ox}}}\right)
\end{equation}
where $Q_{\mathrm{max}} = q N_{\mathrm{mob}} d_{\mathrm{ox}}$, $N_{\mathrm{mob}}$ is the concentration of mobile charges in the oxide, $d_{\mathrm{ox}}$ and $C_{\mathrm{ox}}$ are oxide thickness and capacitance, respectively. The minus sign takes into account that we consider positive charges which should create a negative $\Delta V_{\mathrm{th}}$ that will have a maximum when all ions accumulate near the channel/oxide interface (i.e. $x\,=\,0$). Finally, this ion-induced $\Delta V_{\mathrm{th}}$ is used for piecewise calculation of the drain current as
\begin{equation}
I_{\mathrm{D}}(V_{\mathrm{G}}, t) = \begin{cases}
I_{\mathrm{min}}  \cdot \exp\left(\dfrac{V_{\mathrm{G,eff}} - V_{\mathrm{th}} - \Delta V_{\mathrm{th}}}{SS}\right) & V_{\mathrm{G,eff}} \leq V_{\mathrm{th}} + \Delta V_{\mathrm{th}} \\\mu_{\mathrm{eff}} \cdot n_{\mathrm{2D}} \cdot q \cdot \dfrac{W}{L}  \cdot V_{\mathrm{D}} & V_{\mathrm{G,eff}} > V_{\mathrm{th}} + \Delta V_{\mathrm{th}}
\end{cases}
\end{equation}
where the carrier density $n_{\mathrm{2D}} = C_{\mathrm{eff}}(V_{\mathrm{G,eff}} - V_{\mathrm{th}} - \Delta V_{\mathrm{th}})/q$ with effective capacitance $C_{\mathrm{eff}}$ accounting for the quantum capacitance effects in the MoS$_2$ channel and the capacitance of interface states. The effective gate voltage $V_{\mathrm{G,eff}} = V_{\mathrm{G}} - Q_{\mathrm{it}}/C_{\mathrm{ox}}$ is used to consider possible screening by interface charges. For simplicity, the subthreshold swing SS, minimum current $I_{\mathrm{min}}$ and carrier mobility $\mu_{\mathrm{eff}}$ were inset as input parameters to the model, and $V_{\mathrm{th}}$ was considered to be 0.5$\,$V above the calculated flatband voltage $V_{\mathrm{FB}}$. The use of this approach allowed us to simulate the double sweep $I_{\mathrm{D}}$-$V_{\mathrm{G}}$ characteristics while considering different $t_{\mathrm{sw}}$. The complete description of our compact model with all parameters can be found in the SI. 

\begin{figure}[!h]
\vspace{0mm}
\begin{minipage} {\textwidth} 
\hspace{1.5cm}
  \includegraphics[width=13cm]{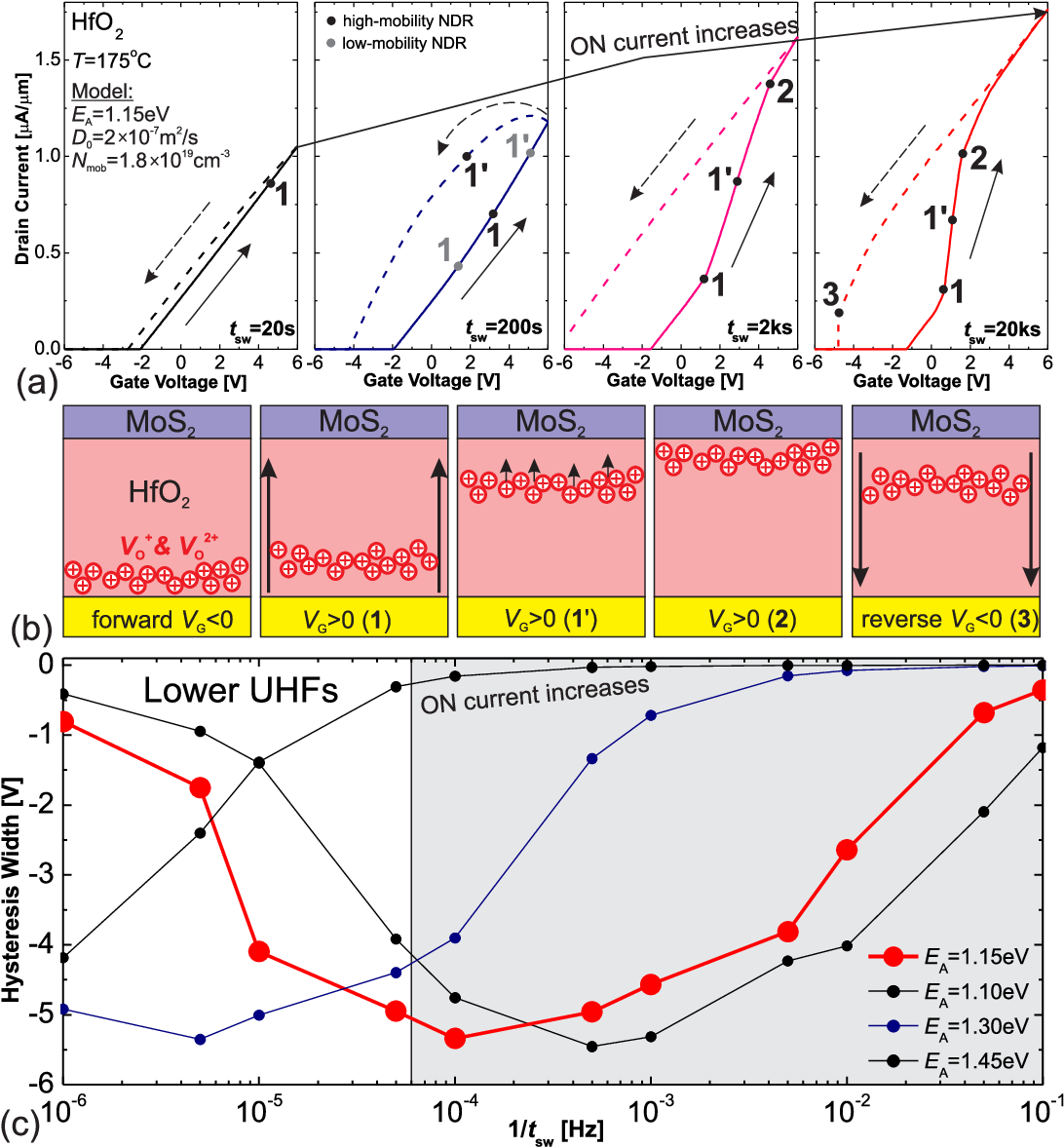} 
\caption{\label{Fig.4} (a) Double sweep $I_{\mathrm{D}}$-$V_{\mathrm{G}}$ characteristics of the MoS$_2$/HfO$_2$ FET simulated with our compact model using different $t_{\mathrm{sw}}$ and $T\,=\,$175\textdegree C. The trends are qualitatively similar to our experimental results shown in Fig.3c. (b) Schematics that illustrate the drift of positive oxygen vacancies in the oxide. At negative $V_{\mathrm{G}}$ they are concentrated at the gate side of HfO$_2$. Then when $V_{\mathrm{G}}$ becomes positive bulk motion towards the MoS$_2$ channel starts if $t_{\mathrm{sw}}$ is slow enough (state 1). However, when approaching MoS$_2$ the charges start feeling the interface effects and thus slowing down (state 1'), before finally getting trapped at the channel side (state 2). The dynamics of the CCW hysteresis and its side features such as the NDR effect and $I_{\mathrm{ON}}$ increase are determined by the relative positions of the points 1, 1' and 2 on the $I_{\mathrm{D}}$-$V_{\mathrm{G}}$ curves that depend on $t_{\mathrm{sw}}$. Also, if the sweep is too slow, the drift of mobile charges back to the gate may start (state 3) when $V_{\mathrm{G}}$ becomes negative during the reverse sweep. (c) The lower UHFs extracted by mapping method from the series of $I_{\mathrm{D}}$-$V_{\mathrm{G}}$ characteristics simulated with different $E_{\mathrm{A}}$. While a maximum of the CCW hysteresis similar to Fig.3b is revealed, a larger $E_{\mathrm{A}}$ shifts it to the slower sweep frequencies and vice versa. The gray box marks typical measurement range. }
\end{minipage}
\end{figure}

We suggest that the CCW hysteresis dynamics that we observe for the MoS$_2$/HfO$_2$ devices originate from the drift of positive oxygen vacancies V$_{\mathrm{O}}^{+}$ and V$_{\mathrm{O}}^{2+}$ in HfO$_2$. Then we have identified that the results of Fig.3 can be qualitatively reproduced with the model by using $E_{\mathrm{A}}\,=\,1.15\,$eV which is close to the literature values for oxygen vacancies in HfO$_2$~\cite{BERSUKER11}, and also physically feasible $N_{\mathrm{mob}}\,=\,3\,\times10^{19}\,$cm$^{-3}$ and $D_0\,=\,2\times 10^{-7}\,$m$^2$/s. The simulated double sweep $I_{\mathrm{D}}$-$V_{\mathrm{G}}$ characteristics which exhibit NDR behavior and the ON current increase are shown in Fig.4a. These results could be interpreted based on the schematics provided in Fig.4b. Namely, at negative $V_{\mathrm{G}}$ positive mobile charges are concentrated at the back gate side of HfO$_2$. As $V_{\mathrm{G}}$ becomes positive, they start moving closer to the MoS$_2$ channel (state 1). Then if $t_{\mathrm{sw}}$ is too short for the ions to respond they are far from the channel and thus the CCW hysteresis is small. However, for slower $t_{\mathrm{sw}}$ the CCW dynamics further develop, as the ions approach the channel making the CCW hysteresis larger, and at the same time start slowing down due to the interface effects (state 1'). In our compact model this is considered by reducing the rate of mobile charges near interfaces (see equation (5) in the SI). Then if $V_{\mathrm{Gmax}}$ is within the free bulk motion of ions, i.e. between the states 1 and 1', high-mobility NDR memory effects appear. This is because ions continue moving towards the channel fast when sweep direction changes. As a result $I_{\mathrm{D}}$ increases against the $V_{\mathrm{G}}$ change as the cumulative $\Delta V_{\mathrm{th}}$ given by equation (2) becomes more negative. Then the pre-condition for NDR is that mobile charges movement lags behind the $V_{\mathrm{G}}$ change. This can be realized also if the state 1' is just before $V_{\mathrm{Gmax}}$ as the ions are still moving though much slower, i.e. low-mobility NDR. However, if $t_{\mathrm{sw}}$ is increased further, mobile charges finally reach the MoS$_2$/HfO$_2$ interface (state 2) and get trapped there. Then the states 1, 1' and 2 appear already during the forward sweep that causes an abrupt kink of $I_{\mathrm{D}}$ instead of NDR. This also results in a sizable increase of $I_{\mathrm{ON}}$ due to self-doping with positive charges. However, since at the state 2 all mobile charges are already trapped at the channel side of oxide, they cannot move anymore even if $V_{\mathrm{G}}$ increases and thus the CCW hysteresis starts to localize below the point 2. Finally, if the sweep is extremely slow, the state 3 followed by a kink downwards appears at a negative $V_{\mathrm{G}}$ during the reverse sweep. This indicates the drift of mobile ions back to the gate side of HfO$_2$. Furthermore, as illustrated by our additional simulation results (see Fig.S4 in the SI), the NDR effect is not present if $N_{\mathrm{mob}}$ is decreased and becomes more pronounced if it is increased.  

In Fig.4c we show the lower UHFs extracted from the $I_{\mathrm{D}}$-$V_{\mathrm{G}}$ curves simulated using different $E_{\mathrm{A}}$. Just like in Fig.3b,e, a distinct maximum of the CCW hysteresis is present at the sweep frequency roughly corresponding to $1/\tau_{\mathrm{drift}}$ that is required for ions to cross $d_{\mathrm{ox}}$, i.e. if state 2 is reached at $V_{\mathrm{Gmax}}$. At the same time, larger migration barrier obviously shifts the maximum to the slower frequencies and vice versa. In Fig.S5 in the SI we also show that if the pre-factor of the diffusion coefficient $D_{\mathrm{0}}$ is made smaller, the maximum also moves to slower frequencies since the ions become slower and need more time to cross the oxide thickness. Based on these observations it is clear that the same hysteresis dynamics could be obtained with different combinations of the three key parameters (see Fig.S6 in the SI). This would obviously make it complicated to use precise fits of the experimental data for extraction of ion parameters. However, we can consider that $E_{\mathrm{A}}$ values which we are using provide a reasonable estimate since any change for more than 0.2$\,$eV would either result is a very different dynamics as compared to our experimental observations, or in non-physical values of $N_{\mathrm{mob}}$ and $D_{0}$. 

\subsubsection{{Boosting the NDR magnitude with narrower sweep ranges}}

Being equipped with fundamental understanding of NDR behavior in our MoS$_2$/HfO$_2$ FETs from the compact model, we next target to confine these memory dynamics by adjusting the $V_{\mathrm{G}}$ sweep range. In Fig.5a we show the double sweep $I_{\mathrm{D}}$-$V_{\mathrm{G}}$ characteristics of the MoS$_2$/HfO$_2$ FET measured at $T\,=\,$175\textdegree C using slowest achieved $t_{\mathrm{sw}}$ and different $V_{\mathrm{G}}$ sweep ranges. Remarkably, while for the sweep range of -6 to 6$\,$V the CCW hysteresis starts to localize, for the sweeps with smaller $V_{\mathrm{Gmax}}$ we still observe the NDR effect within this sweep time range. This is a very intuitive observation, since if $V_{\mathrm{Gmax}}$ is smaller mobile charges need more time to cross $d_{\mathrm{ox}}$ and achieve the state 2 shown in Fig.4a,b. The lower UHFs obtained for this and another device provided in Fig.5b indeed confirm that for narrower sweep ranges the maximum of the CCW hysteresis would be reached at considerably slower sweep frequencies. The full hysteresis mapping results for the Device 1 can be found in Fig.S7 in the SI. They particularly illustrate that due to slower drift of mobile charges for the sweep ranges of -6 to 2$\,$V and -6 to 4$\,$V, a purely CW hysteresis caused by charge trapping is still observed for faster sweeps. 

\begin{figure}[!h]
\vspace{0mm}
\begin{minipage} {\textwidth} 
\hspace{1.5cm}
  \includegraphics[width=13cm]{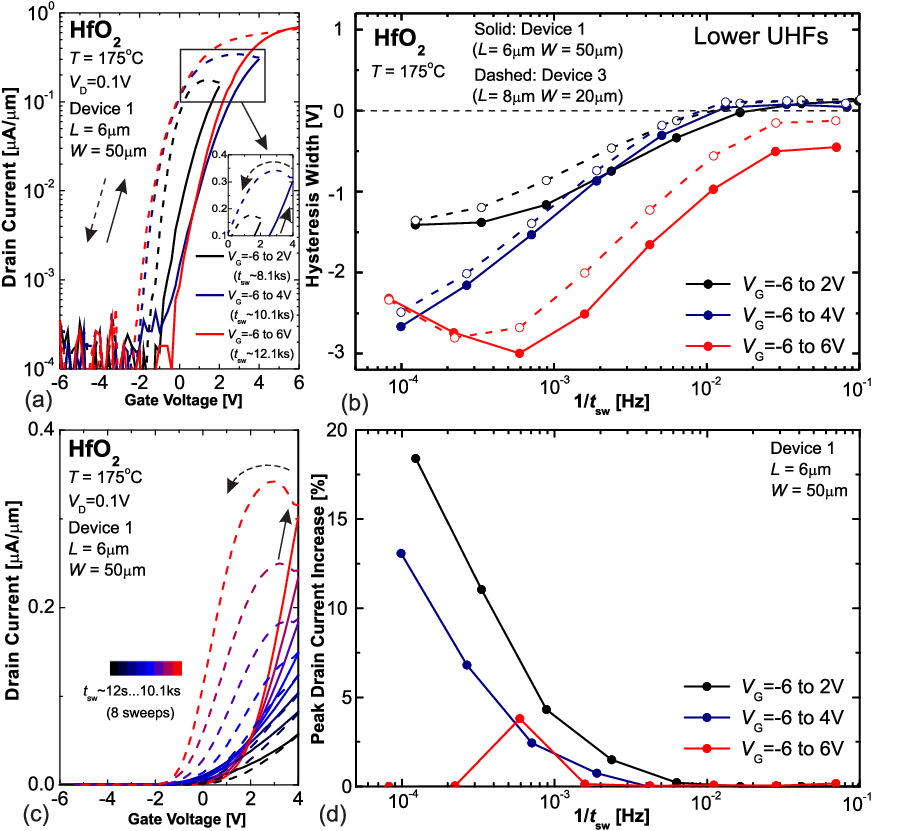} 
\caption{\label{Fig.5} (a) Double sweep $I_{\mathrm{D}}$-$V_{\mathrm{G}}$ characteristics of the MoS$_2$/HfO$_2$ FET measured using slowest achieved $t_{\mathrm{sw}}$ for different $V_{\mathrm{G}}$ sweep ranges and $T\,=\,$175\textdegree C. While for -6 to 6$\,$V the CCW hysteresis already starts to localize, for narrower sweep ranges the NDR effect is still present. (b) The lower UHFs obtained for two devices indeed show that for -6 to 6$\,$V the maximum of CCW hysteresis is reached faster which is because with more positive $V_{\mathrm{Gmax}}$ mobile charges need less time to cross $d_{\mathrm{ox}}$ and thus can localize at the channel side of HfO$_2$ earlier. (c) Full set of double sweep $I_{\mathrm{D}}$-$V_{\mathrm{G}}$ curves measured using -6 to 4$\,$V sweep range in a linear scale. Sizable and progressive NDR effect towards slower sweeps is visible. (d) The dependence of NDR magnitude vs. sweep frequency for different sweep ranges. Narrower sweep ranges reveal stronger effect that appears at slower sweeps.}
\end{minipage}
\end{figure}

In Fig.5c we show the full set of double sweep $I_{\mathrm{D}}$-$V_{\mathrm{G}}$ characteristics measured using -6 to 4$\,$V sweep range in a linear scale. Remarkably, the magnitude of NDR is considerably stronger as compared to the -6 to 6$\,$V sweep range (e.g. Fig.3). In Fig.5d we show the frequency dependence of the NDR magnitude quantified as the peak $I_{\mathrm{D}}$ increase relatively to $I_{\mathrm{D}}$($V_{\mathrm{Gmax}}$) of the forward sweep. Indeed, for narrower sweep ranges NDR is stronger and appears in a broader $t_{\mathrm{sw}}$ range. This returns us back to the compact model interpretation shown in Fig.4a,b. Namely, when using smaller $V_{\mathrm{Gmax}}$ we are dealing with the high-mobility NDR effect that is caused by free motion of mobile charges in the bulk of HfO$_2$. As in that case the ions mostly do not feel the interface effects, the memory-like performance observed within a broad range of $t_{\mathrm{sw}}$. Then its magnitude is mostly defined by the proximity of charges to the channel and obviously increases vs. $t_{\mathrm{sw}}$ until the states 1' and 2 are reached. However, if $V_{\mathrm{Gmax}}$ is more positive, ions pass through the HfO$_2$ too fast and we likely observe the low-mobility NDR effect related to much slower near-interface motion (state 1'). This type of NDR behavior appears only in a narrow $t_{\mathrm{sw}}$ range since at higher $V_{\mathrm{Gmax}}$ the state 2 can be reached too fast, thereby making charges immobile. In Fig.S8 in the SI we demonstrate that the key trends related to the impact of $V_{\mathrm{G}}$ sweep range shown in Fig.5 as well as the temperature can be qualitatively captured by our compact model.  

We note that classification of the observed NDR mechanisms based on our experimental results and compact model may be useful for future design and precise control of memory dynamics in 2D FETs. For instance, the impact of oxide thickness, interface quality and intentional doping with mobile impurities other than preexisting oxygen vacancies could be discovered. 

\subsection{{MoS$_2$/Al$_2$O$_3$ FETs at high temperatures: superior stability}}

\begin{figure}[!h]
\vspace{0mm}
\begin{minipage} {\textwidth} 
\hspace{1.5cm}
  \includegraphics[width=13cm]{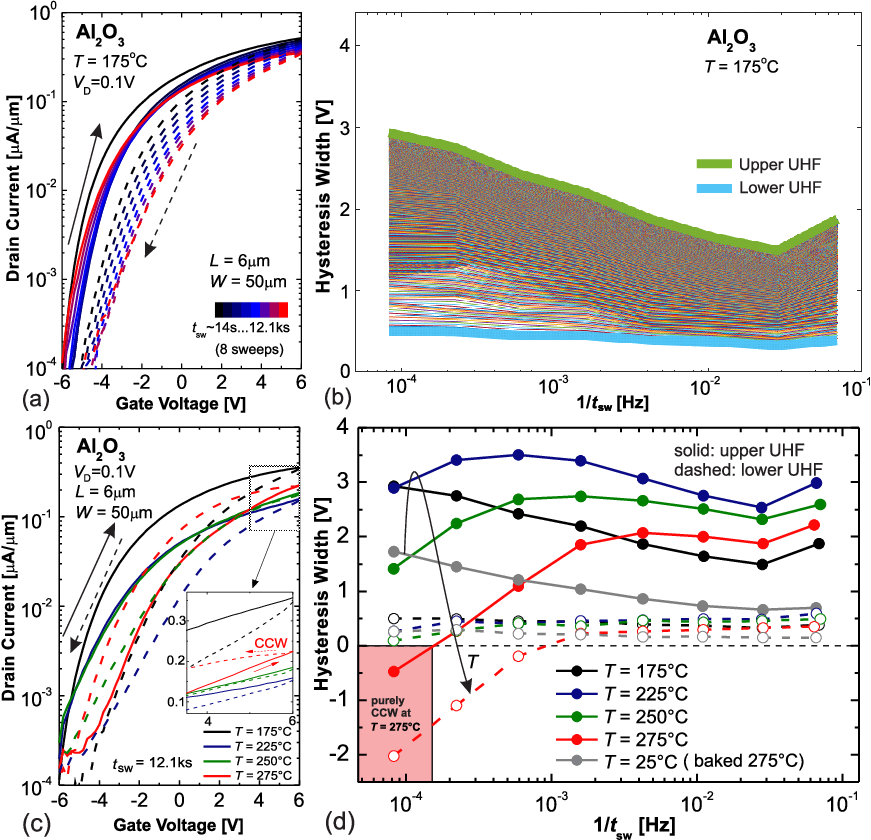} 
\caption{\label{Fig.6} (a) Double sweep $I_{\mathrm{D}}$-$V_{\mathrm{G}}$ characteristics of our MoS$_2$/Al$_2$O$_3$ FET measured at $T\,=\,$175\textdegree C using 8 subsequent sweeps with $t_{\mathrm{sw}}$ up to 12.1$\,$ks. (b) The corresponding mapping results showing that the hysteresis remains purely CW just like it was at room temperature. (c) Double sweep $I_{\mathrm{D}}$-$V_{\mathrm{G}}$ characteristics measured for the same device up to $T\,=\,$225\textdegree C using slowest achieved $t_{\mathrm{sw}}$. Transition to the CCW hysteresis at higher temperatures accompaining with slight self-doping is clearly visible. (d) The corresponding upper and lower UHFs showing change from thermal activation of charge trapping up to $T\,=\,$225\textdegree C to activation of the CCW mechanism related to the drift of oxygen vacancies at higher temperatures. The post-annealing $T\,=\,$25\textdegree C curves that show smaller and purely CW hysteresis nicely match the trends.}
\end{minipage}
\end{figure}

In Fig.6a we show the double sweep $I_{\mathrm{D}}$-$V_{\mathrm{G}}$ characteristics of our MoS$_2$/Al$_2$O$_3$ FET measured at $T\,=\,$175\textdegree C using 8 subsequent sweeps with $t_{\mathrm{sw}}$ up to 12.1$\,$ks. The corresponding hysteresis mapping results provided in Fig.6b confirm that just like it was at room temperature, purely CW hysteresis is present. However, the results measured for the same device up to $T\,=\,$275\textdegree C (Fig.6c,d) still allowed us to catch the interplay between the CW and CCW mechanisms (for better understanding, see the full mapping results for all temperatures in Fig.S9 in the SI). Namely, up to $T\,=\,$225\textdegree C we are dealing with thermally activated charge trapping by oxide traps in Al$_2$O$_3$. This results in a well-known decrease of $I_{\mathrm{D}}$ at slow sweeps (Fig.6c and inset) and a classical bell-shape maximum of the CW hysteresis that shifts to faster frequencies vs. temperature~\cite{ILLARIONOV16A}. Being rarely observed for amorphous oxides with broad defect bands, here this CW maximum is nicely captured within our $t_{\mathrm{sw}}$ range at $T\,=\,$225\textdegree C (blue curve in Fig.6d). However, at $T\,=\,$250\textdegree C thermal activation of mobile charges starts that partially compensates the left part of the CW maximum, and at $T\,=\,$275\textdegree C we finally see a purely CCW hysteresis at slow sweeps that comes together with the current increase due to self-doping, as was discussed above for the devices with HfO$_2$. Remarkably, the $T\,=\,$275\textdegree C results for this MoS$_2$/Al$_2$O$_3$ FET are very similar to those measured for our MoS$_2$/HfO$_2$ device at $T\,=\,$125\textdegree C (see Fig.S3a,b in the SI). This suggests that the origin of the CCW hysteresis in both devices is similar, being related to the drift of oxygen vacancies. However, in Al$_2$O$_3$ the same CCW effects start to appear for the temperatures that are higher by at least 150\textdegree C. Based on our compact model, this suggests that Al$_2$O$_3$ should have $E_{\mathrm{A}}$ for migration of oxygen vacancies of about 1.6$\,$eV as compared to 1.15$\,$eV that we assumed for HfO$_2$. This may be because ionic Al-O bonds are stronger as compared to Hf-O bonds. Therefore, these results clearly show that Al$_2$O$_3$ enables far better stability with respect to the CCW hysteresis caused by drift of oxygen vacancies as compared to HfO$_2$, though at the same time being less relevant to achieve NDR memory performance. 

\subsection{{Bias stress stability of Al$_2$O$_3$ and HfO$_2$ at high temperatures}}

\begin{figure}[!h]
\hspace{0cm}
\vspace{-3mm}
\begin{minipage} {\textwidth} 
\hspace{1cm}
  \includegraphics[width=14cm]{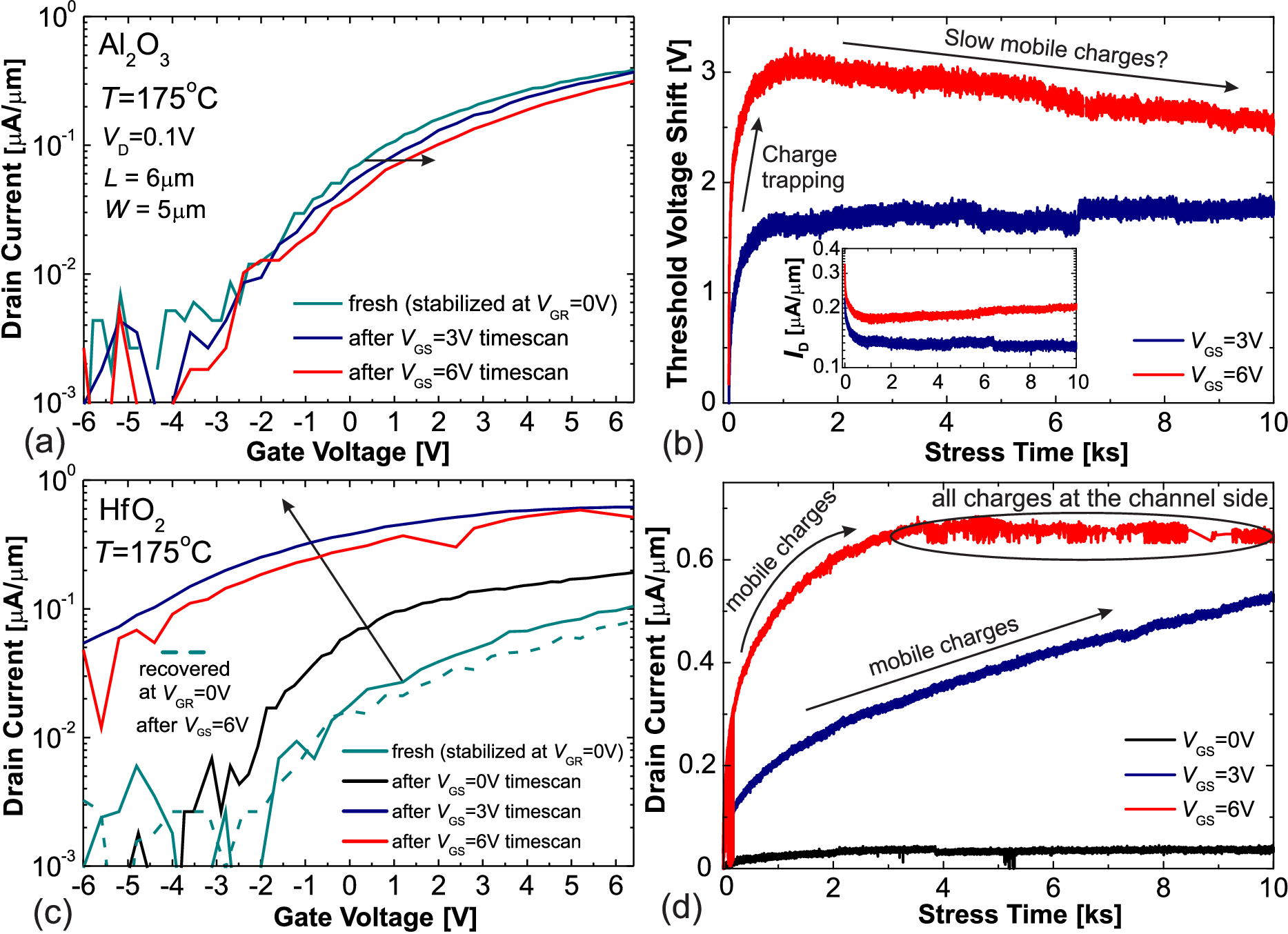} 
\caption{\label{Fig.7} (a) Reference $I_{\mathrm{D}}$-$V_{\mathrm{G}}$ characteristics of our MoS$_2$/Al$_2$O$_3$ FET measured before and after $t_{\mathrm{s}}\,=\,$10$\,$ks gate bias stresses. (b) The $\Delta V_{\mathrm{th}}$ vs. $t_{\mathrm{s}}$ dependences recalculated from the $I_{\mathrm{D}}$ traces shown in the inset. An increase of positive $\Delta V_{\mathrm{th}}$ caused by charge trapping is observed, with possible minor signs of reversal for $V_{\mathrm{GS}}\,=\,$6$\,$V starting from 1$\,$ks that should be due to activation of slow oxygen vacancies in Al$_2$O$_3$. (c) Reference $I_{\mathrm{D}}$-$V_{\mathrm{G}}$ characteristics of the MoS$_2$/HfO$_2$ FETs measured before and after $t_{\mathrm{s}}\,=\,$10$\,$ks gate bias stresses reveal sizable negative shift of $\Delta V_{\mathrm{th}}$ that is due to the drift of positive charges to the channel side of HfO$_2$. (d) The $I_{\mathrm{D}}$ vs. $t_{\mathrm{s}}$ traces clearly show a strong current increase, with possible saturation for $V_{\mathrm{GS}}\,=\,$6$\,$V when all mobile vacancies reach the channel side of HfO$_2$. Extraction of $\Delta V_{\mathrm{th}}$ is barely possible in this case, though its negative sign is obvious. } 
\vspace*{1cm}
\end{minipage}
\end{figure}

To solidify the above findings obtained using hysteresis analysis, we finally examine our MoS$_2$/Al$_2$O$_3$ and MoS$_2$/HfO$_2$ FETs under constant bias stress at $T\,=\,$175\textdegree C. These tests focused on tracking the time evolution of $I_{\mathrm{D}}$ up to $t_{\mathrm{s}}\,=\,10\,$ks at subsequently increased positive stress voltages $V_{\mathrm{GS}}$ following 10$\,$ks stabilization at $V_{\mathrm{GR}}\,=\,0\,$V, while doing fast control $I_{\mathrm{D}}$-$V_{\mathrm{G}}$ sweeps before and after each stress. The results for our MoS$_2$/Al$_2$O$_3$ devices are shown in Fig.7a,b. Using the fresh reference $I_{\mathrm{D}}$-$V_{\mathrm{G}}$ curve obtained after initial device stabilization (Fig.7a), we can recalculate the measured $I_{\mathrm{D}}$($t$) traces (Fig.7b, inset) into the $\Delta V_{\mathrm{th}}$($t$) dependences (Fig.7b). These results confirm that generally we are dealing with positive $\Delta V_{\mathrm{th}}$ that originates from charge trapping by oxide traps and cause a decrease of $I_{\mathrm{D}}$. However, the curves obtained using $V_{\mathrm{GS}}\,=\,6\,$V indicate a clear reversal in the trend with some increase of $I_{\mathrm{D}}$ and thus partial compensation of $\Delta V_{\mathrm{th}}$ starting at about 1$\,$ks. This suggests activation of slow mobile oxygen vacancies in Al$_2$O$_3$. As was discussed above, they should have larger $E_{\mathrm{A}}$ as compared to the same defects in HfO$_2$ and thus cannot be captured in hysteresis measurements as the CCW trends would have appeared at inaccessibly slow $t_{\mathrm{sw}}$. Indeed, as illustrated in Fig.6 they introduce the CCW contribution of hysteresis at higher temperatures. We also note that the $\Delta V_{\mathrm{th}}$ values measured using a constant current mode (Fig.7b) are considerably larger as compared to the ones which we could get by comparing the reference curves (Fig.7a). This suggests that the degradation partially recovers already during few seconds of the control $I_{\mathrm{D}}$-$V_{\mathrm{G}}$ sweeps that is typical for oxide traps at high temperatures~\cite{ILLARIONOV14B}.  

The related results for our MoS$_2$/HfO$_2$ FETs provided in Fig.7c,d show a totally different behavior which matches the CCW hysteresis dynamics discussed above. Already from the reference curves (Fig.7c) we see large negative $\Delta V_{\mathrm{th}}$ after positive bias stresses that comes together with overall increase of $I_{\mathrm{D}}$. The $I_{\mathrm{D}}$($t$) traces (Fig.7d) also show a strong current increase that becomes faster and reaches saturation with $V_{\mathrm{GS}}\,=\,6\,$V. The latter indicates that all charges have reached their equilibrium positions at the channel side of HfO$_2$ (i.e. state 2 in Fig.4b). However, no recalculation into the $\Delta V_{\mathrm{th}}$($t$) dependences is possible since the vertical drifts and transformation of the shapes of $I_{\mathrm{D}}$-$V_{\mathrm{G}}$ curves make any precise definition of $\Delta V_{\mathrm{th}}$ impossible. Still in Fig.S10 in the SI we demonstrate that the obtained $I_{\mathrm{D}}$($t$) traces can be qualitatively captured using our compact model for mobile charges with the parameters similar to those used for hysteresis. Furthermore, observations of Fig.7c such as degradation of SS in the post-stress curves and complete recovery after 10$\,$ks at $V_{\mathrm{GR}}\,=\,0\,$V nicely complement our findings. The former effect can be explained by remote scattering at the positive charges coming closer to the MoS$_2$/HfO$_2$ interface, and the latter one should be due to return of mobile charges back to the gate side of HfO$_2$.          

\section{Conclusions}

In summary, we have performed a detailed comparison of hysteresis dynamics and bias stress stability in MoS$_2$/HfO$_2$ and MoS$_2$/Al$_2$O$_3$ FETs fabricated using the same process up to 275\textdegree C. Our initial findings indicate that room temperature stability limitations for both device types originate from oxide traps that cause CW hysteresis, with Al$_2$O$_3$ showing a slight advantage due to a more favorable fundamental defect band alignment. However, our major  results demonstrate that at higher temperatures the devices with HfO$_2$ exhibit additional severe instabilities which appear as the CCW hysteresis and negative $V_{\mathrm{th}}$ drifts under a positive gate bias. This behavior and accompanying $I_{\mathrm{D}}$ increase with possible NDR features can be nicely described by our compact model for mobile charges. The positive oxygen vacancies (V$_{\mathrm{O}}^{+}$ or V$_{\mathrm{O}}^{2+}$) in HfO$_2$ having migration barriers of about 1$\,$eV are suggested as most likely candidates. Since in Al$_2$O$_3$ these vacancies are expected to be slower due stronger ionic bonds, for our MoS$_2$/Al$_2$O$_3$ FETs the same effect starts to be pronounced in hysteresis measurements only at 275\textdegree C. Our results reveal an essential insulators selection paradigm for the development and integration of 2D FET technology: while Al$_2$O$_3$ is superior to suppress negative $V_{\mathrm{th}}$ drifts for high-temperature logic applications, their HfO$_2$ counterparts can serve as functional active layers that leverage these instabilities to enable intrinsic memory functionality.

\section{Methods}

\textit{Device fabrication} 

MoS$_2$/HfO$_2$ and MoS$_2$/Al$_2$O$_3$ FETs were fabricated using an identical process employing CVD-grown single-layer MoS$_2$ films taken from the same batch. Their fabrication process was arranged as follows. First, the Si/SiO$_2$ substrates were ultrasonically cleaned in acetone and alcohol for 10 minutes each to remove surface contaminants. Local back-gate patterns were then defined via photolithography on the pretreated Si/SiO$_2$ substrates, followed by deposition of a Ni(10$\,$nm)/Au(30$\,$nm) metal stacks as the back-gate electrode using e-beam evaporation. Next, a $\sim$20$\,$nm-thick insulator (HfO$_2$ or Al$_2$O$_3$) was grown via atomic layer deposition (ALD) under optimized process parameters (substrate temperature of 250\textdegree C), guaranteeing the homogeneity and interfacial flatness of the insulator. Subsequently, the MoS$_2$ films were precisely transferred onto the insulator surface via a PMMA-assisted wet transfer method. Residual polymers were removed by acetone and alcohol soaking followed by N$_2$ blow-drying. The MoS$_2$ channels were patterned via photolithography and reactive ion etching. Finally, the source/drain contact regions were precisely defined by photolithography. Ni(10$\,$nm)/Au(30$\,$nm) was deposited via e-beam evaporation followed by a standard lift-off process, thus completing the back-gate device fabrication. 

\textit{Electrical characterization} 
 
Electrical characterization of MoS$_2$/HfO$_2$ and MoS$_2$/Al$_2$O$_3$ FETs was conducted in a vacuum probe station (HCP-O-2, TIANHENG KEYI (SUZHOU) OPTOELEC TECH CO., LTD) with a base pressure of $\sim5\times10^{-6}\ \text{torr}$. All measurements were performed in complete darkness over a temperature range from $25^\circ\mathrm{C}$ to $175^\circ\mathrm{C}$. For electrical measurements we used Keithley 4200A-SCS semiconductor parameter analyzer controlled by a lab-built Python graphical interface to enable uninterrupted long-term testing. Hysteresis dynamics were characterized via double-sweep $I_{\mathrm{D}}$-$V_{\mathrm{G}}$ measurements with a voltage step of 0.2$\,$V. The sweep time $t_{\mathrm{sw}}$ was varied from a few seconds to about 12$\,$ks, and the gate voltage sweep ranges included -6 to 2$\,$V, -6 to 4$\,$V and -6 to 6$\,$V. The hysteresis width was extracted using our universal hysteresis mapping method~\cite{LV25M} and represented using upper and lower UHFs vs. 1/$t_{\mathrm{sw}}$. The bias stress analysis was performed by measuring $I_{\mathrm{D}}$ vs. $t_{\mathrm{s}}$ traces up to 10$\,$ks using $V_{\mathrm{GS}}$ gradually increased from 0 to 6$\,$V, while applying 10$\,$ks stabilizing rounds with $V_{\mathrm{GR}}\,=\,0\,$V and doing control $I_{\mathrm{D}}$-$V_{\mathrm{G}}$ sweeps before each stressing round. Using the initially measured reference $I_{\mathrm{D}}$-$V_{\mathrm{G}}$ curve, we converted these results into $\Delta V_{\mathrm{th}}$($t_{\mathrm{s}}$) dependences for the MoS$_2$/Al$_2$O$_3$ FETs and concluded that it is not possible for the MoS$_2$/HfO$_2$ devices due to non-parallel drifts caused by $I_{\mathrm{D}}$ increase.  

\textit{Compact model} 

Our compact model describes thermally activated hopping of mobile charges in the oxide with the diffusion coefficient $D = D_0 \exp\left(-\frac{q E_{\mathrm{A}}}{k_{\mathrm{B}} T}\right)$ defined by the activation barrier $E_{\mathrm{A}}$ and constant pre-factor $D_0$. Using this diffusion coefficient, we obtain the ion mobility from the Einstein relation and subsequently use it to get their drift velocity under applied gate bias. Next we set the equilibrium positions of positive mobile charges at the gate and channel sides of the oxide for negative and positive $V_{\mathrm{G}}$, respectively, and apply an iterative scheme to calculate the time-dependent positions of ions $x$($t$) during the sweeps. Finally, knowing $x$($t$) we calculate the resulting $\Delta V_{\mathrm{th}}$ caused by the drift of mobile charges, and use it for calculation of $I_{\mathrm{D}}$-$V_{\mathrm{G}}$ curves for different $t_{\mathrm{sw}}$. The model setup is implemented in Matlab. Full description with all equations used can be found in the SI.      

\begin{acknowledgement}
This work is funded by the National Natural Science Foundation of China (W2432040, 62374155, U25A20496), Guangdong Basic and Applied Basic Research Foundation (2024A1515010179, 2023ZT10X010) and Shenzhen Science and Technology Program (20231115150611001). We also acknowledge the support of the Instruments Center for Physical Science at the USTC and the USTC Center for Micro- and Nanoscale Research and Fabrication.
\end{acknowledgement}

\section{Author contributions} 
S.K.Z., H.H.C. and Y.H.W. contributed equally via arranging all measurements and preparing the first draft of the manuscript. Y.F.M. and R.C.Y. fabricated the devices under supervision of Y.Y.S. Y.Z.L. and J.M.H. contributed to hysteresis mapping analysis and data interpretation. Y.Y.I. edited the final manuscript draft, performed compact modeling and supervised the research on electrical characterization.

\section{Competing interests} 
The authors declare no competing interests.


\providecommand{\latin}[1]{#1}
\makeatletter
\providecommand{\doi}
  {\begingroup\let\do\@makeother\dospecials
  \catcode`\{=1 \catcode`\}=2 \doi@aux}
\providecommand{\doi@aux}[1]{\endgroup\texttt{#1}}
\makeatother
\providecommand*\mcitethebibliography{\thebibliography}
\csname @ifundefined\endcsname{endmcitethebibliography}
  {\let\endmcitethebibliography\endthebibliography}{}

\clearpage
\section*{Supporting Information}
\setcounter{figure}{0}
\makeatletter
\renewcommand{\thefigure}{S\arabic{figure}}
\makeatother
\renewcommand{\figdir}{\figdirSI}

\subsection{Full mapping results for hysteresis measured at 25\textdegree C}

\begin{figure}[!h]
\hspace{0cm}
\vspace{-3mm}
\begin{minipage} {\textwidth} 
\hspace{0.5cm}
  \includegraphics[width=15cm]{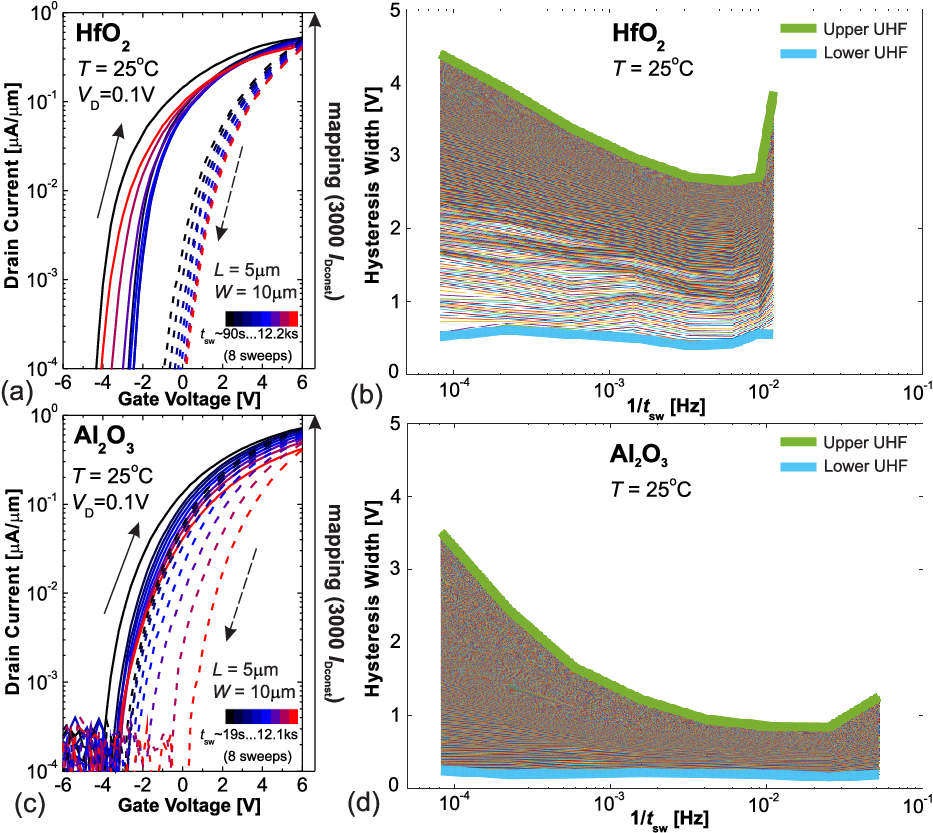} 
\caption{\label{Fig.S1} (a) Double sweep $I_{\mathrm{D}}$-$V_{\mathrm{G}}$ characteristics measured using 8 subsequently increased $t_{\mathrm{sw}}$ for our MoS$_2$/HfO$_2$ FET. (b) The corresponding mapping results showing purely CW hysteresis. (c,d) The related results for a representative MoS$_2$/Al$_2$O$_3$ FET with the same dimensions. A smaller but still purely CW hysteresis is revealed. } 
\vspace*{1cm}
\end{minipage}
\end{figure}

In Fig.S1 we show the full hysteresis mapping results for the representative MoS$_2$/HfO$_2$ and MoS$_2$/Al$_2$O$_3$ FETs with the same channel dimensions. By using the measurement datasets consisting of 8 subsequent $I_{\mathrm{D}}$-$V_{\mathrm{G}}$ sweeps with $t_{\mathrm{sw}}$ of up to about 12$\,$ks (Fig.S1a,c), we scan 3000 constant current points and extract series of $\Delta V_{\mathrm{H}}$(1/$t_{\mathrm{sw}}$) curves. This allows us to obtain the upper and lower UHFs as shown in Fig.S1b,d. Since in these particular cases the hysteresis is purely CW, only the upper UHFs present practical interest for further analysis. At the same time, consistent representation of the final $\Delta V_{\mathrm{H}}$(1/$t_{\mathrm{sw}}$) dependence as the upper UHF is still of key importance since it is far more accurate as compared to the use of just a single randomly selected $I_{\mathrm{Dconst}}$ value. As a result, reasonable comparison of hysteresis dynamics measured for different devices becomes possible. For instance, it is obvious that in our case smaller hysteresis in MoS$_2$/Al$_2$O$_3$ FETs as compared to their MoS$_2$/HfO$_2$ counterparts is related to the device properties rather than extraction methodology.       

\subsection{Additional results for the temperature dependence of hysteresis in MoS$_2$/HfO$_2$ FETs}

\begin{figure}[!h]
\vspace{0mm}
\begin{minipage} {\textwidth} 
\hspace{1cm}
  \includegraphics[width=14cm]{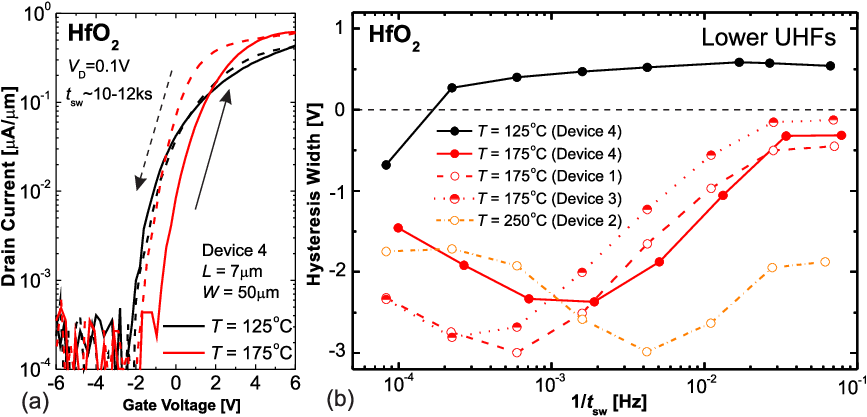} 
\caption{\label{Fig.S2} (a) Double sweep $I_{\mathrm{D}}$-$V_{\mathrm{G}}$ characteristics of the same MoS$_2$/HfO$_2$ FET measured at $T\,=\,$125\textdegree C and 175\textdegree C using slow sweeps. At 125\textdegree C only minor signs of the CCW hysteresis appear since mobile charges are too slow. (b) The corresponding lower UHFs (solid lines) confirm strong thermal activation of mobile charges at 175\textdegree C and 250\textdegree C. This compensates the CW hysteresis coming from charge trapping that is still dominant at 125\textdegree C. The 175\textdegree C curves for two additional devices (dashed and dotted lines) illustrate that the trends in CCW hysteresis are well reproducible.}
\end{minipage}
\end{figure} 

\begin{figure}[!h]
\hspace{0cm}
\vspace{-3mm}
\begin{minipage} {\textwidth} 
\hspace{1.5cm}
  \includegraphics[width=13.5cm]{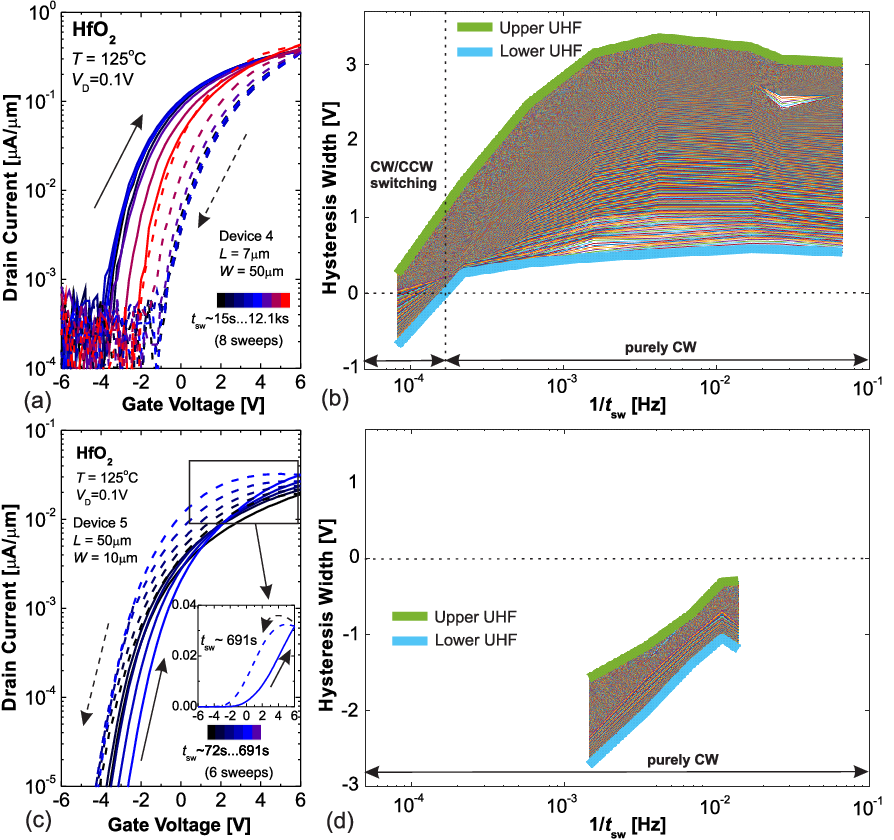} 
\caption{\label{Fig.S3} (a)  Double sweep $I_{\mathrm{D}}$-$V_{\mathrm{G}}$ characteristics of our MoS$_2$/HfO$_2$ FET measured at $T\,=\,$125\textdegree C using 8 subsequent sweeps with $t_{\mathrm{sw}}$ up to 12.1$\,$ks. (b) Full hysteresis mapping results showing the dominant CW hysteresis with some CW/CCW switching only at slow sweeps. (c,d) More fragmentary results for another device measured at $T\,=\,$125\textdegree C 3 months earlier which show purely CCW hysteresis with NDR at relatively fast sweeps.} 
\vspace*{1cm}
\end{minipage}
\end{figure}

In Fig.S2a we show the double sweep $I_{\mathrm{D}}$-$V_{\mathrm{G}}$ characteristics of the same MoS$_2$/HfO$_2$ FET measured at $T\,=\,125\textdegree C and 175\textdegree C using slow sweeps. As confirmed by the corresponding lower UHFs in Fig.S2b, at $T\,=\,125\textdegree C the CW hysteresis is still dominant, with only some CW/CCW switching coming into play at about 10$\,$ks sweep time. This indicates that mobile oxygen vacancies are still slow. However, 175\textdegree C results for 3 devices showing a distinct maximum of the CCW hysteresis clearly illustrate that this temperature is sufficient to activate mobile charges, and at 250\textdegree C the drift of mobile charges is accelerated further.

In Fig.S3a,b we show the full set of $I_{\mathrm{D}}$-$V_{\mathrm{G}}$ characteristics and the corresponding mapping results measured for the Device 4 from Fig.S2 at $T\,=\,125\textdegree C. As the mobile charges are much slower than at 175\textdegree C, the CW hysteresis caused by charge trapping is dominant and some CW/CCW switching appears only at very slow sweeps. However, another device measured 3 months earlier also at $T\,=\,125\textdegree C (Fig.S3c,d) shows purely CCW hysteresis with NDR effects at relatively fast sweeps. According to the modeling results provided in this work, this suggests that the concentration of mobile charges $N_{\mathrm{mob}}$ in that device was much higher. However, subsequent storage of the device could result in partial substitution of oxygen vacancies so that in later measurements the same effects with CCW hysteresis become dominant only at higher temperatures. At the same time, some variability in $N_{\mathrm{mob}}$ due to different local quality of HfO$_2$ may be another reason for the observed difference.   

\subsection{Compact model for the CCW hysteresis caused by positive mobile charges}

The core physical approach of our compact model is to link the ion mobility in the oxide and the threshold voltage of a MoS$_2$ FET through a position-dependent electrostatics. This is done by using the following key steps which are implemented into our Matlab script. 

\textit{\textbf{Step 1: Incorporate thermally activated hopping of mobile charges}} 

The diffusion coefficient for thermally activated hopping of mobile charges is given as 

\begin{equation}
D = D_0 \exp\left(-\frac{q E_{\mathrm{A}}}{k_{\mathrm{B}} T}\right)
\end{equation}

where $D_0$ is a pre-factor coefficient and $E_{\mathrm{A}}$ is the energy activation barrier for the migration of ions. Both these parameters play the key role in defining the dynamics of mobile charges in the oxide. 

\textit{\textbf{Step 2: Obtain the mobility of ions in the oxide from the Einstein relation}}

The Einstein relation links the diffusion coefficient and mobility of a charged particle. Then we can get the following equation for mobility of ions in the oxide

\begin{equation}
\mu = \frac{qD}{k_{\mathrm{B}} T} = \frac{q}{k_{\mathrm{B}} T} \left[D_0 \exp\left(-\frac{q E_{\mathrm{A}}}{k_{\mathrm{B}} T}\right)\right]
\end{equation}
This makes it possible to convert the random diffusion process into directed drift under electric field.

\textit{\textbf{Step 3: Obtain drift velocity of mobile charges under gate bias}}

Knowing the mobility, we can obtain the drift velocity of ions under gate bias which can be represented as

\begin{equation}
v_{\mathrm{drift}} = \mu |E_{\mathrm{ox}}| = \mu \left|\frac{V_{\mathrm{G}}}{d_{\mathrm{ox}}}\right|
\end{equation}
where the oxide electric field $E_{\mathrm{ox}} = V_{\mathrm{G}}/d_{\mathrm{ox}}$ provides the driving force for ion motion.

\textit{\textbf{Step 4: Determine the time evolution of the position of ions}}

Ions move to their equilibrium positions ($x_{\mathrm{eq}}$) which is set by the electric field direction in the oxide. Then we consider that under positive $V_{\mathrm{G}}$ the equilibrium will be near the channel interface ($x_{\mathrm{eq}}\,=\,0.1\,d_{\mathrm{ox}}$), and under negative $V_{\mathrm{G}}$ it will be near the gate electrode ($x_{eq} = 0.9 d_{ox}$). In that case the actual rate of position change towards equilibrium can be determined as 

\begin{equation}
\frac{dx}{dt} = \frac{x_{\mathrm{eq}} - x}{\tau_{\mathrm{drift}}} 
\end{equation}

with the characteristic drift time $\tau_{\mathrm{drift}} = d_{\mathrm{ox}}/v_{\mathrm{drift}}$ representing the time required for ions to cross the oxide. By discretizing the analytical solution of this differential equation, we can obtain the position update of ions at every next time step as 

\begin{equation}
 dx = (x_{\mathrm{eq}} - x_{\mathrm{current}}) \times \Big[1 - \exp\Big(-B\frac{dt}{\tau_{\mathrm{drift}}}\Big)\Big]
\end{equation}

with the current position $x_{\mathrm{current}}$ being considered as the initial position for each step in the loop which states that $x_{\mathrm{new}}\,=\,x_{\mathrm{current}}\,+\,dx$. Note that considering that the sweeps start at negative $V_{\mathrm{G}}$, for positive charges we use the averaged starting position 0.8$d_{\mathrm{ox}}$ at the first step (i.e. close to the gate). The constant parameter $B$ takes into account slowing of ions near the interfaces and their faster motion in the oxide bulk. In our qualitative model setup we use the value of 0.01 if the normalized position of ion $x/d_{\mathrm{ox}}\,<\,0.15$ (i.e. too close to the channel/oxide interface), 0.03 for $0.15\,<\,x/d_{\mathrm{ox}}\,<\,0.3$, 5 for $0.15\,<\,x/d_{\mathrm{ox}}\,<\,0.85$ and 0.02 for $x/d_{\mathrm{ox}}\,>\,0.85$ (i.e. too close to the gate/oxide interface). The physical meaning of 1/$\tau_{\mathrm{drift}}$ is the relaxation rate which thus can be adjusted via multiplying by the position factor $B$. We note that in principle the diffusion contribution coming from random motion of ions could be added to $dx$. However, here we assume that the drift caused by applied gate bias has the dominant impact. 

Finally, physical constraints $x_{\mathrm{new}} = \max(0.1 d_{\mathrm{ox}}, \min(0.9 d_{\mathrm{ox}}, x_{\mathrm{new}}))$ are set to prevent non-physical positions that are too close to interfaces which ions cannot occupy due to their finite sizes and interface barriers.

\textit{\textbf{Step 5: Calculation of the threshold voltage shift induced by motion of mobile charges}}

Knowing the positions of mobile charges at every time step, we next calculate the time-dependent threshold voltage shift which will be induced by their motion in the oxide. It can be written as 

\begin{equation}
\Delta V_{\mathrm{th}}(t) = -\frac{Q_{\mathrm{max}}}{C_{\mathrm{ox}}} \left(1 - \frac{x(t)}{d_{\mathrm{ox}}}\right)
\end{equation}

where $Q_{\mathrm{max}} = q N_{\mathrm{mob}} d_{\mathrm{ox}}$, $N_{\mathrm{mob}}$ is the concentration of mobile charges in the oxide and $C_{\mathrm{ox}}\,=\,\varepsilon_{\mathrm{ox}}\varepsilon_{0}/d_{\mathrm{ox}}$ is the oxide capacitance. The minus sign takes into account that we consider positive charges which should create a negative $\Delta V_{\mathrm{th}}$ that will have a maximum when all ions accumulate near the channel/oxide interface (i.e. $x\,=\,0$).

\textit{\textbf{Step 6: Calculation of the time-dependent drain current}}

To calculate the drain current using the obtained $\Delta V_{\mathrm{th}}$ induced by the ion drift, we first evaluate the interface charge based on the Fermi-Dirac distribution as 

\begin{equation}
Q_{\mathrm{it}} = q D_{\mathrm{it}} k_{\mathrm{B}} T \ln\left[1 + \exp\left(\frac{q(V_{\mathrm{G}} - V_{\mathrm{FB}})}{k_{\mathrm{B}} T}\right)\right]
\end{equation}
where $D_{\mathrm{it}}$ is the density of interface states at the channel/oxide interface and the flat band voltage is determined as 

\begin{equation}
V_{\mathrm{FB}} = -\phi_{\mathrm{b}} - \frac{E_{\mathrm{G}}}{2} - \frac{q D_{\mathrm{it}} E_{\mathrm{G}}}{2 C_{\mathrm{ox}}}
\end{equation}

with $\phi_{\mathrm{b}}$ being the Schottky barrier height at the source/MoS$_2$ interface and $E_{\mathrm{G}}$ the MoS$_2$ bandgap. 

Then the effective gate voltage which takes into account possible screening by interface charges can be obtained as 

\begin{equation}
V_{\mathrm{G,eff}} = V_{\mathrm{G}} - \frac{Q_{\mathrm{it}}}{C_{\mathrm{ox}}}
\end{equation}

Finally, the ion-induced $\Delta V_{\mathrm{th}}$ calculated above can be used for piecewise calculation of the drain current as
\begin{equation}
I_{\mathrm{D}}(V_{\mathrm{G}}, t) = \begin{cases}
I_{\mathrm{min}}  \cdot  \exp\left(\dfrac{V_{\mathrm{G,eff}} - V_{\mathrm{th}} - \Delta V_{\mathrm{th}}}{SS}\right) & V_{\mathrm{G,eff}} \leq V_{\mathrm{th}} + \Delta V_{\mathrm{th}} \\\mu_{\mathrm{eff}} \cdot n_{\mathrm{2D}} \cdot q \cdot \dfrac{W}{L} \cdot V_{\mathrm{D}} & V_{\mathrm{G,eff}} > V_{\mathrm{th}} + \Delta V_{\mathrm{th}}
\end{cases}
\end{equation}
where the carrier density is $n_{\mathrm{2D}} = C_{\mathrm{eff}}(V_{\mathrm{G,eff}} - V_{\mathrm{th}} - \Delta V_{\mathrm{th}})/q$ with effective capacitance $C_{\mathrm{eff}} = C_{\mathrm{ox}} (C_{\mathrm{q}} + C_{\mathrm{it}})/(C_{\mathrm{ox}} + C_{\mathrm{q}} + C_{\mathrm{it}})$ accounting for the quantum capacitance effects in the MoS$_2$ channel via $C_{\mathrm{q}} = q^2 m_{\mathrm{eff}}/\pi \hbar^2$ and the capacitance of interface states as $C_{\mathrm{it}} = q D_{\mathrm{it}}$ with $m_{\mathrm{eff}}$ being the effective carrier mass in MoS$_2$ and $D_{\mathrm{it}}$ the density of interface states, respectively. The equilibrium threshold voltage $V_{\mathrm{th}}$ is set to be 0.5$\,$V above the calculated $V_{\mathrm{FB}}$ for simplicity.

\textit{\textbf{Implementation and input parameters of the model}}

In our Matlab implementation we set the $V_{\mathrm{G}}$ arrays from $V_{\mathrm{Gmin}}$ to $V_{\mathrm{Gmax}}$ and back and assign the corresponding time point from the time array of 0 to $t_{\mathrm{sw}}$ to each $V_{\mathrm{G}}$ point. This obviously makes the results dependent on the input $t_{\mathrm{sw}}$ and the $V_{\mathrm{Gmin}}$ to $V_{\mathrm{Gmax}}$ sweep range so that the hysteresis dynamics observed experimentally could be reproduced qualitatively.

The input material-related parameters of the MoS$_2$/HfO$_2$ FETs include the insulator thickness $d_{\mathrm{ox}}\,=\,20\,$nm, the permittivity of HfO$_2$ $\varepsilon_{\mathrm{ox}}\,=\,20$, the bandgap of MoS$_2$ $E_{\mathrm{G}}\,=\,2.53\,$eV, the Schottky barrier height $\phi_{\mathrm{b}}\,=\,0.2\,$eV and the effective electron mass in MoS$_2$ $m_{\mathrm{eff}}\,=\,0.45\,m_{0}$. 

For the simplicity, we also input the minimum current in the OFF state $I_{\mathrm{min}}\,=\,10^{-13}\,$A, effective carrier mobility in MoS$_2$ $\mu_{\mathrm{eff}}\,=\,10\,$cm$^2$V$^{-1}$s$^{-1}$ and the subthreshold swing SS$\,=\,280\,$mV/dec to the equation (10). The density of interface states $D_{\mathrm{it}}\,=\,-2\times10^{12}\,$cm$^{-2}$eV$^{-1}$ is used to consider possible acceptor-type fixed charges at the channel/oxide interface. The drain voltage $V_{\mathrm{D}}\,=\,0.1\,$V is set just like in our experiments, as well as the representative channel length $L\,=\,6\,\mu$m and width $W\,=\,50\,\mu$m. 

Finally, the key parameters which determine the dynamics of ion drift in the oxide are the total concentration of mobile charges $N_{\mathrm{mob}}$, the activation energy of their migration $E_{\mathrm{A}}$ and the constant pre-factor $D_0$ which represents the maximum possible diffusion coefficient if the activation barrier is removed. The best qualitative agreement with our experimentally observed hysteresis dynamics is obtained using $N_{\mathrm{mob}}\,=\,3\,\times10^{19}\,$cm$^{-3}$, $E_{\mathrm{A}}\,=\,1.15\,$eV and $D_0\,=\,2\times 10^{-7}\,$m$^2$/s. However, for the proof-of-concept demonstrations we also changed these values to illustrate the key trends as specified in the figures.

\subsection{Impact of $N_{\mathrm{mob}}$ and $D_0$ on the CCW hysteresis dynamics from the compact model}

\begin{figure}[!h]
\hspace{0cm}
\vspace{-3mm}
\begin{minipage} {\textwidth} 
\hspace{2cm}
  \includegraphics[width=11.8cm]{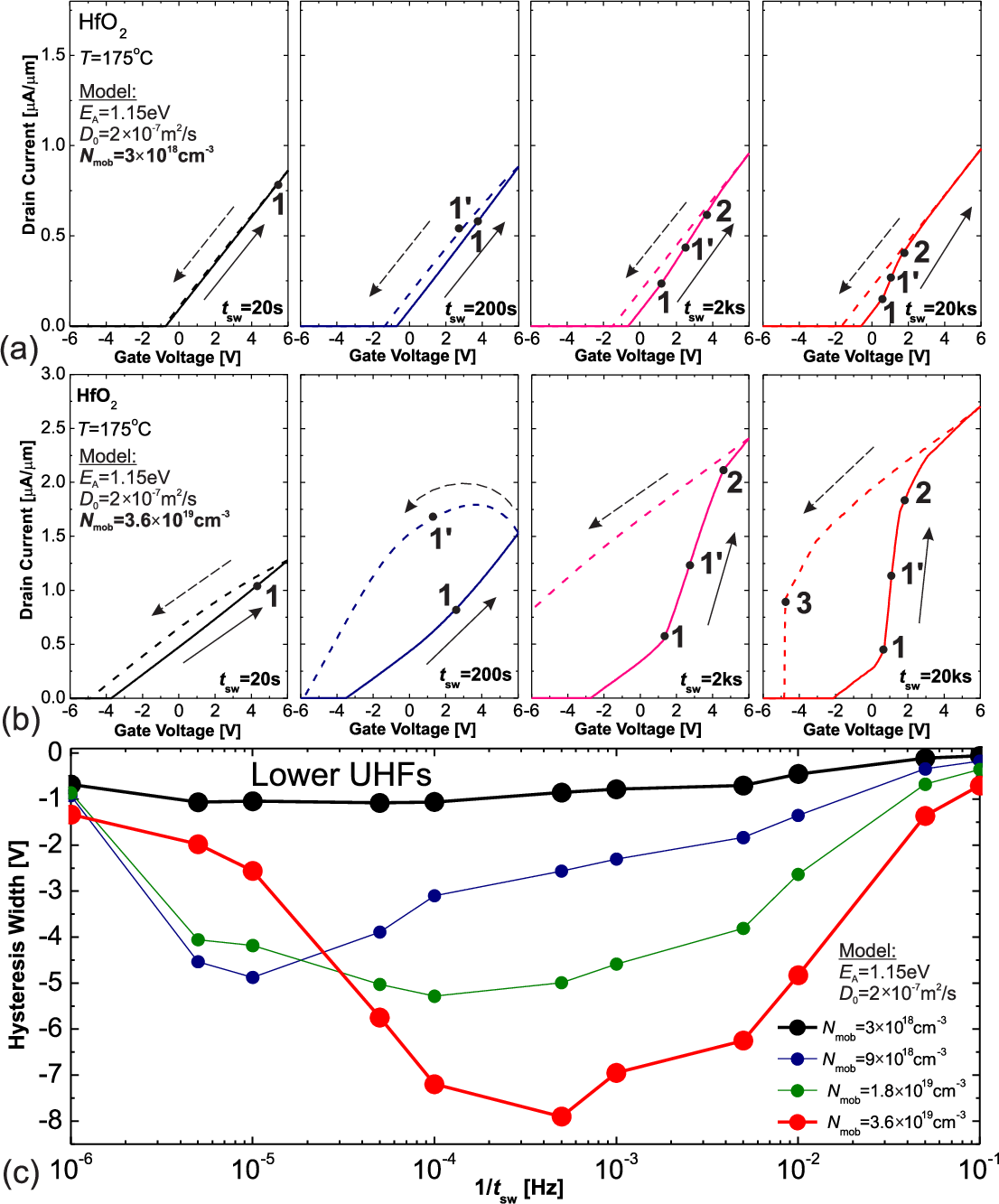} 
\caption{\label{Fig.S4} (a) Double sweep $I_{\mathrm{D}}$-$V_{\mathrm{G}}$ characteristics of the MoS$_2$/HfO$_2$ FET simulated with our compact model considering reduced $N_{\mathrm{mob}}\,=\,3\,\times10^{18}\,$cm$^{-3}$. No NDR effect is observed while an increase in $I_{\mathrm{ON}}$ is minor. (b) The related results obtained using $N_{\mathrm{mob}}\,=\,3.6\,\times10^{19}\,$cm$^{-3}$. Strong NDR effect is present at moderate $t_{\mathrm{sw}}$. (c) Lower UHFs extracted from the simulation results obtained using different $N_{\mathrm{mob}}$. If there are too many mobile charges in the oxide, the maximum of CCW hysteresis is larger and appears at faster sweeps. } 
\vspace*{1cm}
\end{minipage}
\end{figure}

In Fig.S4a we show the double sweep $I_{\mathrm{D}}$-$V_{\mathrm{G}}$ characteristics of the MoS$_2$/HfO$_2$ FET simulated with our compact model considering a reduced $N_{\mathrm{mob}}\,=\,3\,\times10^{18}\,$cm$^{-3}$. For this minor concentration of mobile charges in HfO$_2$, the CCW hysteresis is small while the self-doping is insufficient to cause any NDR effect or sizable increase of the ON current. However, an obvious kink between the points 1 and 2 is still visible for slow sweeps. This indicates that migration of positive charges to the channel side of HfO$_2$ still takes place, even though their concentration is small. In contrast, the results simulated with higher $N_{\mathrm{mob}}\,=\,3.6\,\times10^{19}\,$cm$^{-3}$ (Fig.S4b) reveal a strong NDR effect at moderate sweeps, as well as a more pronounced increase of $I_{\mathrm{ON}}$ which at slow $t_{\mathrm{sw}}$ turns into a large and abrupt kink between the points 1 and 2. In Fig.S4c we show the lower UHFs extracted from the $I_{\mathrm{D}}$-$V_{\mathrm{G}}$ curves simulated using different $N_{\mathrm{mob}}$. It is obvious that for larger concentrations of ions the CCW hysteresis is larger, while the corresponding maximum is more distinct and appears at faster sweep frequencies. 

In Fig.S5 we show that the pre-factor of the diffusion coefficient $D_{\mathrm{0}}$ also affects the dynamics of CCW hysteresis considerably. If $D_{\mathrm{0}}$ is made smaller, the ions obviously become slower and need more time to cross the oxide thickness. This appears in the lower UHFs as a parallel shift of the maximum to slower frequencies. At the same time,  $D_{\mathrm{0}}$ does not affect the shape of the maximum, which is different from the impact of $N_{\mathrm{mob}}$ illustrated in Fig.S4c. 

\begin{figure}[!h]
\hspace{0cm}
\vspace{-3mm}
\begin{minipage} {\textwidth} 
\hspace{2cm}
  \includegraphics[width=11.8cm]{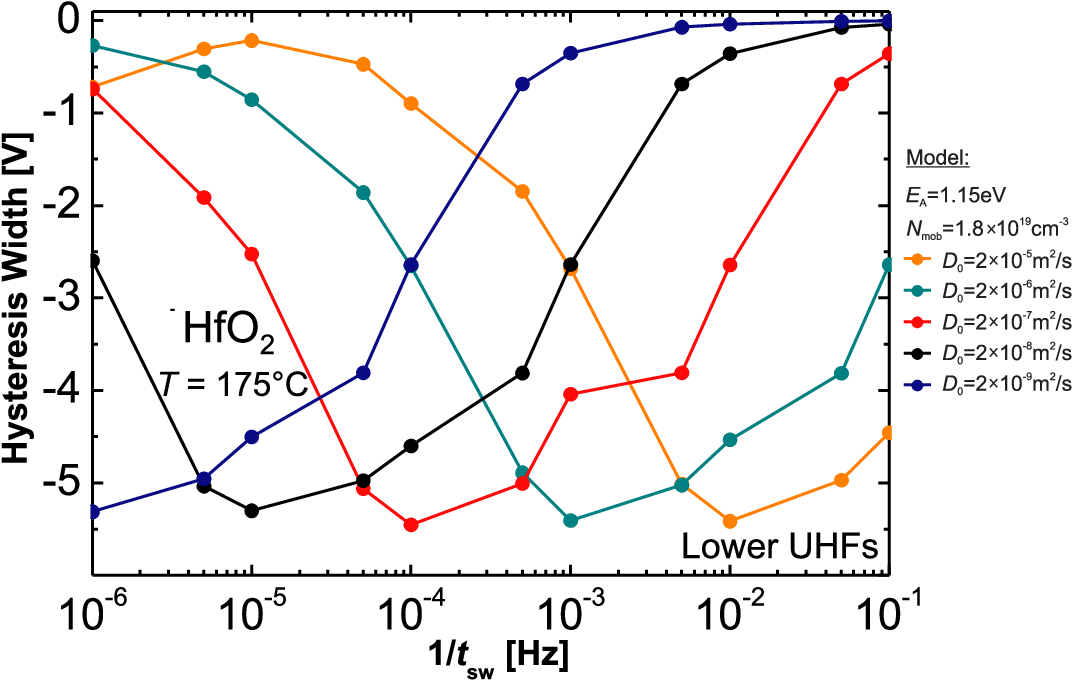} 
\caption{\label{Fig.S5} Lower UHFs extracted from the simulation results for the MoS$_2$/HfO$_2$ FET obtained using different $D_{\mathrm{0}}$. Smaller diffusion pre-factor obviously shifts the maximum of CCW hysteresis to slower sweep frequencies. } 
\vspace*{1cm}
\end{minipage}
\end{figure}

\subsection{Equivalent sets of parameters in the compact model}

The dynamics of mobile ions in HfO$_2$ will be mostly determined by $N_{\mathrm{mob}}$, $E_{\mathrm{A}}$ and $D_0$ which are not known precisely. While our analysis qualitatively reproduces experimental findings using the Set 1 with $N_{\mathrm{mob}}\,=\,3\,\times10^{19}\,$cm$^{-3}$, $E_{\mathrm{A}}\,=\,1.15\,$eV and $D_0\,=\,2\times 10^{-7}\,$m$^2$/s, it is obvious that because of several degrees of freedom the compact model can produce identical hysteresis dynamics also using alternative combinations of parameters. Few examples that would be still physically feasible are listed in Fig.S6a and indeed produce the same hysteresis dynamics as illustrated by the lower UHFs provided in Fig.S6b. 

\begin{figure}[h]
\hspace{0cm}
\vspace{-3mm}
\begin{minipage} {\textwidth} 
\hspace{2.5cm}
  \includegraphics[width=10.8cm]{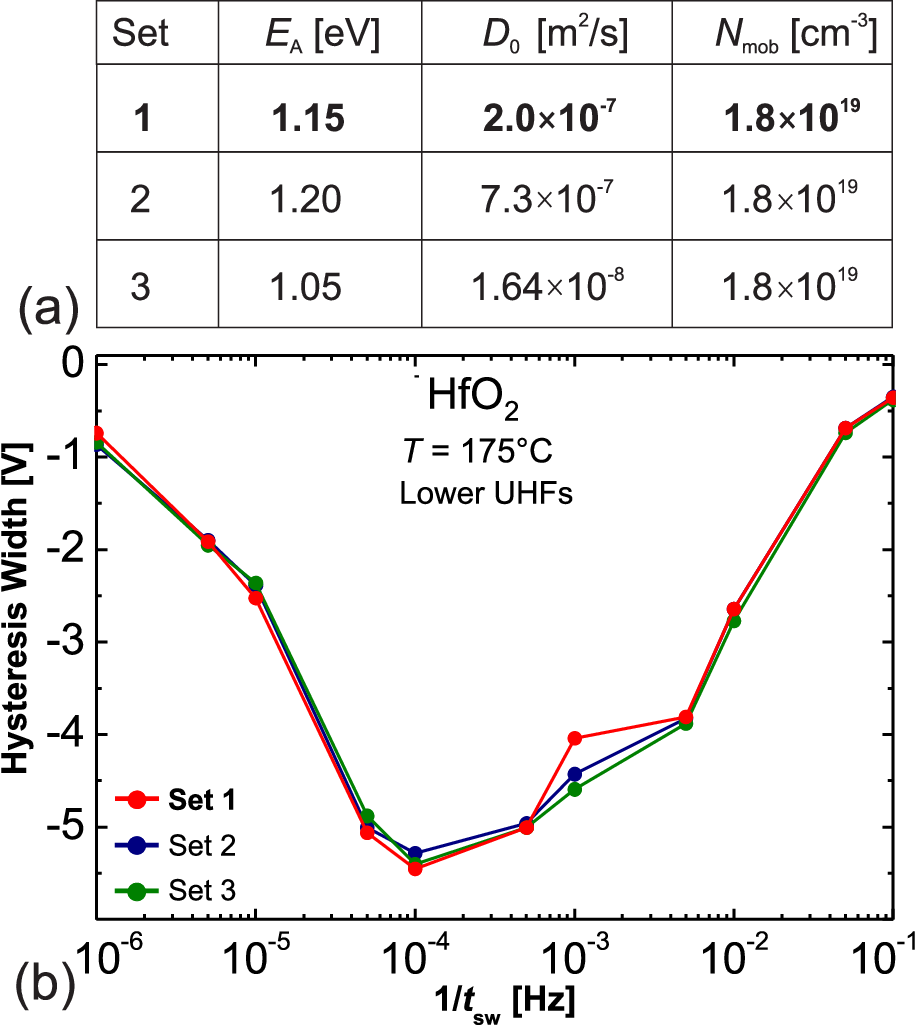} 
\caption{\label{Fig.S6} (a) Examples of physically feasible equivalent sets of the key parameters of mobile charges in HfO$_2$ that would result in identical hysteresis dynamics according to the model. (b) The lower UHFs extracted from the $I_{\mathrm{D}}$-$V_{\mathrm{G}}$ curves simulated using these 3 parameter sets that indeed look the same.} 
\vspace*{1cm}
\end{minipage}
\end{figure}

We see that the model is very sensitive to $E_{\mathrm{A}}$ and even a minor increase of the activation barrier would need a considerably larger $D_0$ (Set 2). Alternatively, a minor decay in $E_{\mathrm{A}}$ would require a much smaller $D_0$ to produce the same CCW hysteresis (Set 3). Therefore, we can conclude that considerable deviation of $E_{\mathrm{A}}$ from the range 1.05-1.25$\,$eV that should be typical for oxygen vacancies in HfO$_2$ would likely result in non-physical values of $N_{\mathrm{mob}}$ and $D_0$.   

\subsection{Full mapping results for the sweep range dependence of the CCW hysteresis in MoS$_2$/HfO$_2$ FETs}

\begin{figure}[!h]
\hspace{0cm}
\vspace{-3mm}
\begin{minipage} {\textwidth} 
\hspace{1.5cm}
  \includegraphics[width=13.5cm]{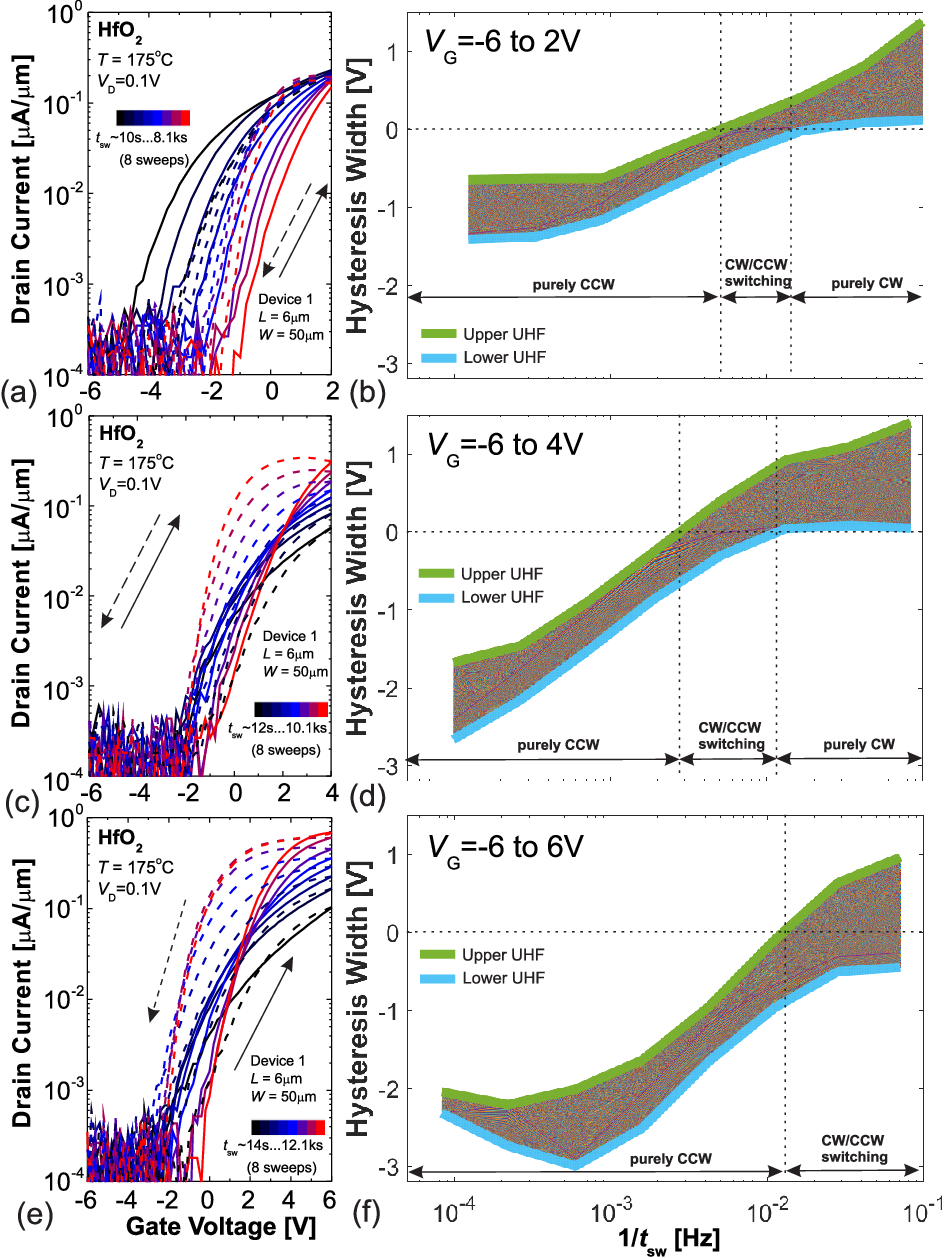} 
\caption{\label{Fig.S7} Double sweep $I_{\mathrm{D}}$-$V_{\mathrm{G}}$ characteristics of our MoS$_2$/HfO$_2$ FET measured at $T\,=\,$175\textdegree C using 8 subsequent sweeps with $t_{\mathrm{sw}}$ up to 8.1, 10.1 or 12.1$\,$ks and full hysteresis mapping results for the sweep ranges of -6 to 2$\,$V (a,b), -6 to 4$\,$V (c,d) and -6 to 6$\,$V (e,f). Frequency ranges with different hysteresis dynamics can be clearly separated based on the signs of the upper and lower UHFs.} 
\vspace*{1cm}
\end{minipage}
\end{figure}

In Fig.S7 we provide the full set of $I_{\mathrm{D}}$-$V_{\mathrm{G}}$ characteristics and the corresponding mapping results measured for our MoS$_2$/HfO$_2$ FET at $T\,=\,$175\textdegree C using different $V_{\mathrm{G}}$ sweep ranges. It is obvious that for narrower sweep ranges NDR effects and CCW hysteresis maximum are observed at considerably slower sweeps. This is because mobile charges in HfO$_2$ need more time to reach the channel side if $V_{\mathrm{Gmax}}$ is smaller. As a result, purely CW hysteresis is present for faster sweep frequencies in the case of -6 to 2$\,$V and -6 to 4$\,$V sweep ranges, thereby indicating that mobile charges are still unable to fully compensate charge trapping. Remarkably, with the full mapping results we can clearly identify and separate the frequency ranges with the purely CW hysteresis, CW/CCW hysteresis switching and purely CCW hysteresis.  

\clearpage
\subsection{Sweep range and temperature dependence of the CCW hysteresis in MoS$_2$/HfO$_2$ FETs: compact model}

\begin{figure}[!h]
\hspace{0cm}
\vspace{-3mm}
\begin{minipage} {\textwidth} 
\hspace{1.5cm}
  \includegraphics[width=13.5cm]{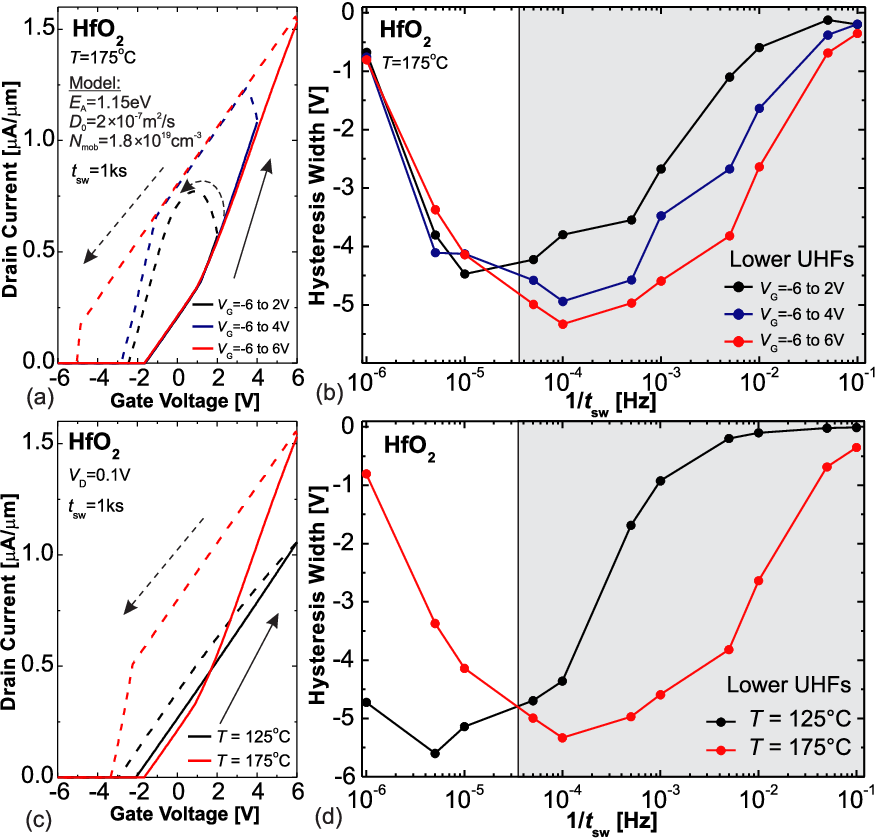} 
\caption{\label{Fig.S8} (a) Double sweep $I_{\mathrm{D}}$-$V_{\mathrm{G}}$ characteristics of the MoS$_2$/HfO$_2$ FET simulated using our compact model while assuming different $V_{\mathrm{G}}$ sweep ranges, $t_{\mathrm{sw}}\,=\,1\,$ks and $T\,=\,$175\textdegree C. Just like in the experiments, for -6 to 2$\,$V we still observe NDR effect but for -6 to 6$\,$V localization of the CCW starts. (b) The corresponding lower UHFs also confirm that acceleration of mobile charges with more positive $V_{\mathrm{Gmax}}$ is well described by the model. (c) Double sweep $I_{\mathrm{D}}$-$V_{\mathrm{G}}$ characteristics of the MoS$_2$/HfO$_2$ FET simulated with the same model parameters for $T\,=\,$125\textdegree C and 175\textdegree C assuming $t_{\mathrm{sw}}\,=\,1\,$ks. At $T\,=\,$175\textdegree C the CCW hysteresis is obviously larger. (d) The corresponding lower UHFs confirm that at higher temperature the CCW hysteresis maximum shifts to faster sweep frequencies.}  
\vspace*{1cm}
\end{minipage}
\end{figure}

In Fig.S8a,b we provide the results obtained using our compact model for different $V_{\mathrm{G}}$ sweep ranges. Faster change of the CCW hysteresis from the NDR to localized behavior for wider sweep ranges observed in the experiments (e.g. Fig.S7) is nicely confirmed by the $I_{\mathrm{D}}$-$V_{\mathrm{G}}$ characteristics shown in Fig.S8a. The corresponding lower UHFs (Fig.S8b) show that for narrower sweep range the maximum of the CCW hysteresis shifts to slower sweep frequencies, which also goes in line with our experimental findings. Finally, in Fig.S8c,d we show the modeling results for $T\,=\,125$\textdegree C and 175\textdegree C. They nicely confirm that the CCW hysteresis maximum shifts to faster sweep frequencies due to thermal activation of mobile charges which we also observe in our experiments. 

\subsection{Full mapping results for hysteresis in MoS$_2$/Al$_2$O$_3$ FETs up to 275\textdegree C}

In Fig.S9 we provide the full set of $I_{\mathrm{D}}$-$V_{\mathrm{G}}$ characteristics and the corresponding mapping results measured for our MoS$_2$/Al$_2$O$_3$ FET at $T\,=\,$175\textdegree C, $T\,=\,$225\textdegree C, $T\,=\,$250\textdegree C and $T\,=\,$275\textdegree C. These results show that up to $T\,=\,$250\textdegree C we are dealing with the purely CW hysteresis which appears as a bell-shape maximum of the upper UHFs that shifts to faster frequencies due to thermal activation. This behavior originates from charge trapping by oxide traps in Al$_2$O$_3$ that also causes a typical transformation of the shape of $I_{\mathrm{D}}$-$V_{\mathrm{G}}$ curves with current decay for slower sweeps. At $T\,=\,$250\textdegree C a certain compensation of the CW hysteresis at slow sweeps may be present, and finally at $T\,=\,$275\textdegree C the slow sweep hysteresis is purely CCW. This reveals activation of mobile oxygen vacancies that comes together with the current increase for slow sweeps.     

\begin{figure}[!h]
\hspace{0cm}
\vspace{-3mm}
\begin{minipage} {\textwidth} 
\hspace{2cm}
  \includegraphics[width=11.0cm]{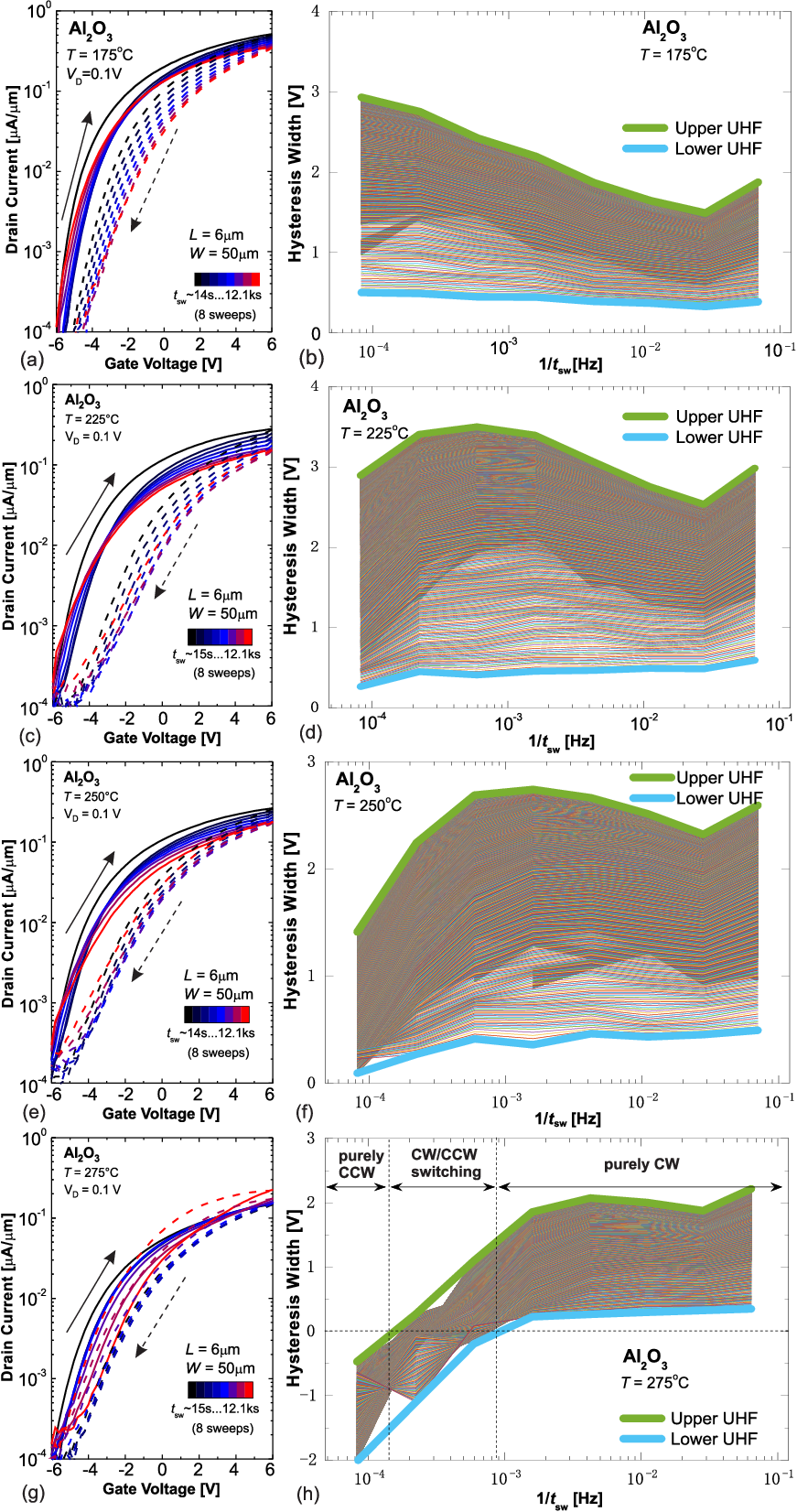} 
\caption{\label{Fig.S9} Double sweep $I_{\mathrm{D}}$-$V_{\mathrm{G}}$ characteristics of our MoS$_2$/Al$_2$O$_3$ FET measured using 8 subsequent sweeps with $t_{\mathrm{sw}}$ up to 12.1$\,$ks and full hysteresis mapping results for $T\,=\,175\textdegree C (a,b); $T\,=\,225\textdegree C (c,d); $T\,=\,250\textdegree C (e,f); $T\,=\,275\textdegree C (g,h).} 
\vspace*{1cm}
\end{minipage}
\end{figure}

\clearpage

\subsection{Qualitative modeling of $I_{\mathrm{D}}$($t$) traces for MoS$_2$/HfO$_2$ FETs}

In Fig.S10 we show the $I_{\mathrm{D}}$($t$) dependences for a MoS$_2$/HfO$_2$ FET at $T\,=\,$175\textdegree C obtained using our compact model for mobile charges with the parameters similar to the ones used for hysteresis. The key trends are the same as in our experiments. Namely, the current increases versus time as positive mobile charges come closer to the channel side of HfO$_2$. At a certain time saturation of $I_{\mathrm{D}}$ takes place when all of them reach their equilibrium positions. This saturation happens faster if $V_{\mathrm{GS}}$ is more positive. However, we note that the initial parts of these traces depend on the starting position of ions input into the simulations, or in fact in their starting distribution in the oxide. Here we consider an averaged value of 0.75$d_{\mathrm{ox}}$.  

\begin{figure}[!h]
\hspace{0cm}
\vspace{-3mm}
\begin{minipage} {\textwidth} 
\hspace{1.5cm}
  \includegraphics[width=13.5cm]{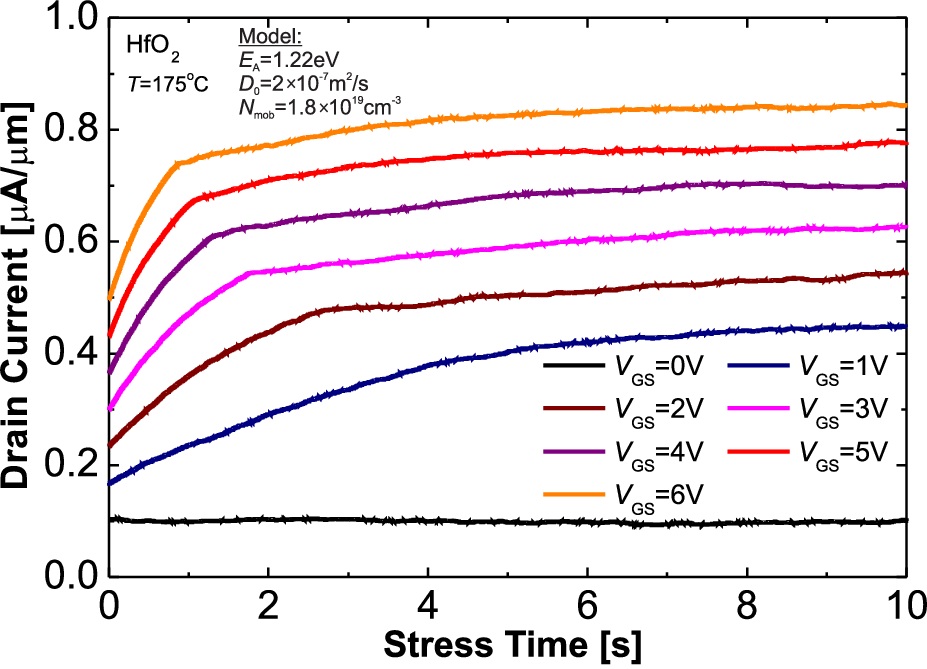} 
\caption{\label{Fig.S10} Simulated $I_{\mathrm{D}}$($t$) dependences for a MoS$_2$/HfO$_2$ FET at $T\,=\,$175\textdegree C. Larger $V_{\mathrm{GS}}$ makes saturation of $I_{\mathrm{D}}$ faster since mobile charges need less time to cross the oxide.} 
\vspace*{1cm}
\end{minipage}
\end{figure}

\end{document}